\documentclass[12pt]{article} 
\pdfoutput=1
\usepackage{graphicx,floatflt,amssymb,rotate}
\usepackage{mathrsfs}
\usepackage{amssymb}
\usepackage{amsmath}
\usepackage{color}
\usepackage{graphicx}
\usepackage{subfigure}
\usepackage{multirow}
\usepackage{float}
\usepackage{pstricks}
\usepackage[toc,page]{appendix}
\usepackage[mathscr]{euscript}
\usepackage{cite}
\usepackage[colorlinks=true,
linkcolor=red,
urlcolor=blue,
citecolor=blue]{hyperref}
\usepackage{color}

\def\mET{E_T \hspace{-1.0em}/\;\:}

\begin{document}
\vskip 30pt 

\begin{center}
	{\Large \bf Revisiting the high-scale validity of Type-II seesaw model with novel LHC signature} \\
		\vspace*{1cm} {\sf ~Dilip Kumar Ghosh$^{a,}$\footnote{tpdkg@iacs.res.in}, 
			~Nivedita Ghosh$^{a,}$\footnote{tpng@iacs.res.in},
			~Ipsita Saha$^{b,}$\footnote{ipsita.saha@roma1.infn.it} and 
                        ~Avirup Shaw$^{a,}$\footnote{avirup.cu@gmail.com}}\\
		\vspace{10pt} {\small } {\em $^a$Department of Theoretical Physics, 
			Indian Association for the Cultivation of Science,\\
			2A $\&$ 2B, Raja S.C. Mullick Road, Kolkata 700032, India \\
			$^b$ Istituto Nazionale di Fisica Nucleare, Sezione di Roma, \\
			Piazzale Aldo Moro 2, I-00185 Roma, Italy}   
\end{center}

\begin{abstract}  
\noindent 
The Type-II seesaw model is a well-motivated new physics scenario
to address the origin of the neutrino mass issue. We show that this model can easily 
accommodate an absolutely stable vacuum until the Planck scale, however
with strong limit on the exotic scalar masses and the corresponding mixing angle. 
We examine the model prediction at the current and future high luminosity run of the Large Hadron Collider (LHC) 
for the scalar masses and mixing angles fixed at such high-scale valid region. 
Specifically, we device the associated and pair production of the charged scalars as a new probe of the model at the LHC. 
We show that for a  particular signal process the model can be tested with $5\sigma$ signal significance even at the present run of the LHC.

\vskip 5pt \noindent  

\end{abstract}

\renewcommand{\thesection}{\Roman{section}}  
\setcounter{footnote}{0}  
\renewcommand{\thefootnote}{\arabic{footnote}}

\section{Introduction}

The discovery of the 125 GeV resonance at the Large Hadron Collider (LHC)~\cite{Aad:2012tfa,Chatrchyan:2012xdj} with its increasing biasness 
towards the Standard Model (SM) values for couplings, anyway, has set forth one of the most discussed questions of this time that asks for an absolute stable electroweak vacuum up to very high scale,
say Grand Unification Scale (GUT) $(10^{16} ~\rm GeV)$ or Planck Scale $(10^{19}~ \rm GeV)$.
The evolution of the Higgs quartic coupling only with 
SM interaction, is unable to maintain its boundedness from below
much before the GUT or Planck scale and becomes negative at around $10^{10} ~\rm GeV$ even after taking into account all the uncertainties in determining the top quark mass and strong coupling constant ~\cite{EliasMiro:2011aa,Degrassi:2012ry,Alekhin:2012py,Bezrukov:2012sa,Masina:2012tz,Buttazzo:2013uya}. This subsequently
calls for intervention of some new physics scenario at or before such scale. 
Although, a {\it metastable} vacuum could be a possibility, the quest for the absolute stability can be advocated for the hunt for new physics (NP) scenarios. To affirm the positivity of the Higgs quartic coupling all its way during the renormalization group (RG) evolution, positive aid from extra scalar bosons are mandatory to
negotiate the negative fermionic pull that comes dominantly from the top quark.
Therefore, NP models that extends the SM scalar sector can easily serve the purpose. In view of this, the triplet scalar extension, alias, the Type-II seesaw model has already been proven a well-deserved NP candidate to ameliorate the vacuum stability problem~\cite{Chun:2012jw,Chao:2012mx,Dev:2013ff,Haba:2016zbu,Dev:2017ouk}. 
In order to seek an absolutely stable vacuum in the 
Type-II seesaw regime, one must also maintain the perturbative unitarity of all the
new couplings introduced lately. It has been studied that all these 
stability and perturbative unitarity requirements together
can dramatically control the allowed region of the model parameter space in view of the
non-standard scalar masses and mixing even only at the electroweak (EW) scale~\cite{Chabab:2015nel,Das:2016bir}. 
Nevertheless, a complete picture of the allowed parameter space in light
of high-scale validity of this scenario in terms of the 
physical masses and mixing has not been emphasized before.
  
On a different note, the Type-II seesaw model~\cite{Schechter:1980gr,Magg:1980ut,Cheng:1980qt,Lazarides:1980nt,Mohapatra:1980yp, Lindner:2016bgg}
has mainly been preferred for its 
attributes towards the generation of tiny nonzero neutrino masses through the
seesaw mechanism. Unlike the other two variants of seesaw, Type-II does
not inherit an extended fermion sector, rather it includes an extra
$SU(2)_L$ triplet scalar to the SM particle content and the lepton number
is broken by two units through a trilinear coupling $(\mu)$ in the scalar potential. The neutrinos acquire Majorana mass term through the Yukawa interaction of these triplet
with the lepton doublets and the small neutrino masses are proportional to 
the product of the Yukawa couplings $(Y_\Delta)$ and the trilinear coupling $(\sim {Y_\Delta \mu v_d^2}/{M_{\Delta}^2})$ 
where, $M_\Delta$ is the triplet scalar mass parameter and $v_d$ is the SM vacuum expectation value (vev). 
The smallness of the neutrino mass can thus be translated by the trilinear coupling 
$\mu$, which is protected by symmetry and can be small following the {\it t'hooft's}
naturalness criterion \cite{tHooft:1979rat}. In fact, this $\mu$ term is responsible for
the vev of the triplet ($v_t$) 
as it yields the tadpole term for the scalar triplet 
due to the spontaneous electroweak symmetry breaking (EWSB) as $v_t \sim ({\mu v_d^2}/{M_\Delta^2})$. 
In general, a proper tuning between
these two parameters ($\mu$ or $v_t$ and $Y_\Delta$) in accordance with the
current neutrino oscillation data \cite{deSalas:2017kay} can generate the tiny neutrino masses even 
keeping the newly introduced scalar masses 
around the EW scale $(M_\Delta \sim v_d)$ unlike 
the other two types of seesaw mechanisms where the new fermion masses
necessarily has large values ($10^9-10^{10} ~\rm GeV$). 
Therefore, the exotic triplet
scalars of Type-II seesaw models lie in the reach of the current collider experiments which makes this 
particular seesaw type to be phenomenologically more appealing.
In addition, the presence of both singly and doubly charged scalars make this
model very attractive from the point of view of exotic particles search at the colliders.

In this work, we rekindle the idea of probing the 
Type-II seesaw model parameter space at 
the LHC that are conclusively allowed by all the stability and 
perturbativite unitarity requirements until the Planck scale. 
The search strategy could thus be restrictive as the extra theoretical constraints 
have put stringent bound on the non standard scalar mass
splitting. The splitting between the singly and doubly charged scalar masses is also
tuned by the one-loop T-parameter constraint at the EW scale \cite{Chun:2012jw,Baak:2014ora}. It is therefore expected 
that the allowed parameter region in the high-scale valid space will be more
contrived. Moreover, to have the lightest CP-even state as the 125 GeV resonance state
with the SM-Higgs like coupling, one must ensure that the signal strengths for various
decay modes do not overshoot the current experimental limit. In this 
scenario only the loop induced decay mode of 
the Higgs boson, $h \to \gamma \gamma $\cite{Arhrib:2011vc,Carena:2012xa} 
can be affected due to the presence of extra singly and doubly 
charged scalars and hence it is important to consider the limit on the corresponding
signal strength ($\mu_{\gamma \gamma}$) as an additional constrain.

The triplet vev $(v_t)$ plays a crucial role in this model. 
The value of $v_t$ is highly constrained by the $\rho$-parameter and 
quantitatively it can not be larger than a few GeV \cite{Aoki:2012jj}.
This has a greater implication on the decay 
modes of the singly and doubly charged scalars. To be specific, for $v_t < 10^{-4} ~\rm GeV$, the triplet scalars dominantly decay  into leptonic final state and 
for $v_t > 10^{-4} ~\rm GeV$, only gauge boson final states or cascade decays of charged scalars
(if they are kinematically allowed)
are possible~\cite{Perez:2008ha,Melfo:2011nx,Aoki:2011pz}. The latest
same-sign dilepton searches at the LHC  
have already put strong lower limit on doubly charged scalar mass ($>$ 770 - 800 GeV)~\cite{ATLAS:2017iqw}.
However, this limit is only valid for $v_t < 10^{-4}~\rm GeV$
when the doubly charged scalar decays to dilepton pair with $100\%$ branching ratio.
The collider search becomes more involved for $v_t\sim 1~\rm GeV$, due to more complicated decay patterns of the singly and doubly charged scalars. As a result, the lower bounds on these scalars are very relaxed.
Some collider analyses for this case have been done before ~\cite{Akeroyd:2005gt, Akeroyd:2009hb, Akeroyd:2010je, Akeroyd:2011zza, Chiang:2012dk} and after \cite{Kanemura:2013vxa, kang:2014jia, Kanemura:2014goa,Kanemura:2014ipa,Chen:2014qda,Han:2015hba, Han:2015sca, Mitra:2016wpr} the Higgs discovery but our analysis goes beyond what has been reported so far.
In our analysis, we demonstrate the complete picture of a high-scale valid Type-II seesaw potential and
analyze the signal significance at the LHC in the allowed parameter space. 
To be specific, for collider study, we pursue the two different final state topologies,
 $3\ell^\pm +\mET$ and  $2 \ell^\pm + 4j+ \mET$, where, $\ell \equiv e, \mu $.
 The first channel ($3\ell^\pm +\mET$) comes from the associate production of the singly and doubly charged scalars, while the second channel ($2 \ell^\pm + 4j+ \mET$) gets contributions from 
 the associated production as well as from the doubly charged scalar pair production process.
Performing a cut-based analysis to reduce the SM background efficiently,
we explore the possibility of probing the model parameter space in the 
current and high luminosity run of the 13 TeV LHC. 
Such comprehensive study in view of simultaneous consideration of high scale stability and collider prediction 
is completely new.

The paper is organised in the following way. In section II we discuss the model and set our conventions. Then we briefly discuss the various theoretical and experimental constraints applied on 
the model parameter space in section III. Following this, in section IV we describe the features of the model emerged from high scale stability requirement. 
Next, in section V we perform the collider analysis. 
Finally, we conclude in section VI.

\section{$Y=2$, Higgs Triplet Model in a nutshell}\label{2}
In this section, we briefly discuss the Type-II seesaw model.  
It contains an $SU(2)_L $ triplet scalar field $\Delta$ with hypercharge $Y=2$ in addition to the SM fields. 
\begin{eqnarray}
\Delta &=& \frac{\sigma^i}{\sqrt 2}\Delta_i = \left(\begin{array}{cc}
\delta^+/\sqrt 2 & \delta^{++}\\
\delta^0 & -\delta^+/\sqrt 2
\end{array}
\right),
\end{eqnarray}

where $\Delta_1=(\delta^{++}+\delta^0)/\sqrt 2,~\Delta_2=i(\delta^{++}-\delta^0)/\sqrt 2,~\Delta_3=\delta^+$. The complete Lagrangian of this scenario 
is given by: 
\begin{eqnarray}
{\cal L} = {\cal L}_{\rm Yukawa}+{\cal L}_{\rm Kinetic}-V(\Phi,\Delta),
\label{lag}
\end{eqnarray}
where the kinetic and Yukawa interactions are respectively \cite{Arhrib:2011uy}
\begin{eqnarray}
\label{y2kinetic}
{\cal L}_{\rm kinetic} &=&\left(D_\mu\Phi\right)^\dag \left(D^\mu\Phi\right) +{\rm Tr}\left[\left(D_\mu \Delta\right)^\dag \left(D^\mu\Delta\right)\right], \\
\label{y2Yukawa}
{\cal L}_{\rm Yukawa} &=& {\cal L}_{\rm Yukawa}^{\rm SM}- \left(Y_\Delta\right)_{ij} L_i^{\sf T}Ci\sigma_2\Delta L_j+{\rm h.c.}\, .
\end{eqnarray}

Here $\Phi^{\sf T}=(\phi^+ ~~~\phi^0)$ is the SM scalar doublet and $L$ represents 
$SU(2)_L$ left-handed lepton doublet. Neutrino Yukawa coupling is represented 
by $Y_\Delta$, and $C$ is the Dirac charge conjugation matrix. 
Covariant derivative of the scalar triplet field is given by:
\begin{equation} 
D_\mu \Delta = \partial_\mu \Delta + i\frac{g}{2}[\sigma^a W_\mu^a,\Delta]
+i g^\prime B_\mu \Delta \qquad (a=1,2,3).
\end{equation}
$\sigma^a$ are the Pauli matrices, $g$ and $g^\prime$ are the gauge coupling constants of the $SU(2)_L$ and $U(1)_Y$ group respectively.

The most general scalar potential is given as \cite{Arhrib:2011uy}:
\begin{eqnarray} 
V(\Phi,\Delta) &=& -m^2_\Phi(\Phi^\dag \Phi)+\frac{\lambda }{4}(\Phi^\dag \Phi)^2+M^2_\Delta {\rm Tr}(\Delta ^\dag \Delta)+ \left(\mu \Phi^{\sf T}i\sigma_2\Delta^\dag\Phi+{\rm h.c.}\right)+\,\nonumber\\
&& \lambda _1(\Phi^\dag\Phi){\rm Tr}(\Delta ^\dag \Delta)+\lambda _2\left[{\rm Tr}(\Delta ^\dag \Delta)\right]^2+\lambda _3{\rm Tr}(\Delta ^\dag \Delta)^2+\lambda _4\Phi^\dag\Delta\Delta^\dag\Phi.
\label{eq:Vpd}
\end{eqnarray}

All the parameters of the potential can be taken to be real 
without loss of generality. After the EWSB,
the minimization of the potential calculates the two mass parameters as,
\begin{equation}
m_\Phi^2 = \lambda \frac{v_d^2}{4} - \sqrt{2} \mu v_t + \frac{(\lambda_1+\lambda_4)}{2}v_t^2 \,,
\label{eq:minim1}
\end{equation}
\begin{equation}
M_\Delta^2 = \frac{\mu v_d^2}{\sqrt{2}v_t} - \frac{\lambda_1 + \lambda_4}{2}v_d^2 - (\lambda_2 + \lambda_3)v_t^2 \,,
\label{eq:minim}
\end{equation}

where, $v_d$ and $v_t$ stands for the doublet and triplet vev respectively
and the electroweak vev is given by $v = \sqrt{v_d^2 + 2v_t^2} = 246 ~\rm GeV$.
The triplet vev $(v_t)$ contributes to the electroweak gauge 
boson masses $M_W^2$ and $M_Z^2$ at tree level, 
$M_W^2 = \frac{g^2(v^2_d+2 v^2_t)}{4}$ and
$M_Z^2 = \frac{g^2(v^2_d+4 v^2_t)}{4 \cos^2 \theta_W}$ 
respectively\footnote{$\theta_W $ is the Weinberg angle.},
thus modifying the SM $\rho$-parameter as:
\begin{eqnarray}\label{y2rho}
\rho = \frac{M_W^2}{M_Z^2\cos^2\theta_W} = \frac{1+\frac{2v_t^2}{v_d^2}}{1+\frac{4v_t^2}{v_d^2}}\,.
\end{eqnarray}
The electroweak precision data constraints require the 
$\rho$-parameter to be very close to its SM value of unity 
and from the latest data: $\rho = 1.0004^{+0.0003}_{-0.0004}$
\cite{Olive:2016xmw}. Consequently, one gets an
upper bound on $\frac{v_t}{v_d} < 0.02$ or $v_t < 5 $ GeV. 
Hence, the triplet vev $v_t$ remains much smaller than the doublet
vev $v_d$.  
On the other hand, from minimization condition one gets 
$v_t \propto \frac{\mu v_d^2}{M_{\Delta}^2}$ which further contributes 
to the neutrino mass generation as $M_\nu = \sqrt{2}{v_t Y_\Delta}$.
The mass matrix $M_\nu$ is diagonalized by the neutrino mixing matrix,
i.e. Pontecorvo-Maki-Nakagawa-Sakata (PMNS) matrix, 
the components of which are determined by the neutrino oscillation data
for a particular neutrino mass hierarchy (for further details, see ~\cite{Dev:2013ff,Bonilla:2015jdf}).  
Therefore, neutrino mass ${\cal O}(0.1)$ eV can be obtained by tuning
either the triplet vev $(v_t)$
or the Yukawa coupling $(Y_\Delta)$.
For $v_t \sim {\cal O} (~\rm GeV)$, the Yukawa coupling has to be 
small, while an order one neutrino Yukawa coupling $(Y_\Delta \sim {\cal O} (1))$, 
demands $v_t \sim {\cal O}(10^{-2}~\rm eV)$. This sets the two extreme limits of the triplet vev.

After EWSB, the scalar fields expanded around respective vevs, can be parametrized as 
\begin{eqnarray}
\Phi &=& \frac{1}{\sqrt{2}}\left(\begin{array}{c}
\sqrt{2}\chi_d^+ \\
v_d + h_d + i\eta_d \\
\end{array}\right) \qquad 
\Delta = \frac{1}{\sqrt 2}\left(\begin{array}{cc}
\delta^+ & \sqrt{2}\delta^{++}\\
v_t + h_t + i\eta_t & -\delta^+
\end{array}
\right)\;. 
\end{eqnarray}
As a consequence, the scalar spectrum contains seven physical Higgs bosons: two doubly charged 
$H^{\pm\pm}$, two singly charged $H^\pm$, two CP-even neural ($h,H$) and a 
CP-odd ($A$) Higgs particles. The mass matrix diagonalizations are done using orthogonal rotation 
matrices comprising of rotation angles $\alpha, \beta $ and $\beta^\prime$ 
respectively for the CP-even, CP-odd and charged sector.

The corresponding mixing angles are given as 
\begin{subequations}
\begin{eqnarray}
\tan\beta' &=& \frac{\sqrt 2 v_t}{v_d},\label{mix1} \quad
\tan\beta =\frac{2v_t}{v_d}\equiv \sqrt 2 \tan\beta'\label{mix2}\\
{\rm and }~~
\tan{2 \alpha} &=& \frac{2\mathcal{B}}{\mathcal{A}-\mathcal{C}}, \\
{\rm where,} \quad
\mathcal{A} = \frac{\lambda }{2} v_d^2,&&
\mathcal{B} = v_d[-\sqrt{2}\mu+(\lambda _1+\lambda _4)v_t],\quad
\mathcal{C} = \frac{\sqrt{2}\mu v^2_d+4(\lambda _2+\lambda _3)v^3_t}{2v_t}. \label{mix3}
\end{eqnarray} 
\end{subequations}

From the scalar potential of Eq.~(\ref{eq:Vpd}) it is evident that there lies eight independent parameters.
Among which the two bilinear terms $m_\Phi^2$ and $M_\Delta^2$ can be traded off for the two vevs $(v_d,v_t)$ 
using the minimization conditions given in Eqs.~(\ref{eq:minim1}) and (\ref{eq:minim}). The remaining five scalar quartic coupling 
and the lepton number violating parameter $\mu$ can be expressed in terms of the five scalar masses and the neutral
scalar mixing angle $\alpha$, for convenience~\cite{Arhrib:2011uy}.
\begin{subequations}
\begin{eqnarray}
\lambda &=& \frac{2}{v_d^2}( c_\alpha^2 m_h^2 +  s_\alpha^2 m_H^2)\,, \label{eq:lambda}\\
\lambda_1 &=& \frac{4 m_{H^\pm}^2}{v_d^2 + 2 v_t^2} - \frac{2 m_{A}^2}{v_d^2 + 4 v_t^2} + \frac{\sin 2\alpha}{2 v_d v_t}(m_h^2 - m_H^2)\,, \label{eq:lambda1}\\
\lambda_2 &=& \frac{1}{v_t^2}\left[\frac{1}{2}\left(s_\alpha^2 m_h^2 + c_\alpha^2 m_H^2\right) + 
\frac{1}{2}\frac{v_d^2 m_A^2}{v_d^2 + 4 v_t^2} - \frac{2 v_d^2 m_{H^\pm}^2}{v_d^2 + 2v_t^2} + m_{H^{\pm\pm}}^2\right]\,, \label{eq:lambda2}\\
\lambda_3 &=&  \frac{1}{v_t^2}\left[ \frac{2 v_d^2 m_{H^\pm}^2}{v_d^2+2v_t^2} - m_{H^{\pm\pm}}^2 - \frac{v_d^2m_A^2}{v_d^2 + 4v_t^2} \right]\,, \label{eq:lambda3} \\
\lambda_4 &=& \frac{4 m_A^2}{v_d^2 + 4v_t^2} - \frac{4 m_{H^\pm}^2}{v_d^2 + 2 v_t^2}\,, \label{eq:lambda4}\\
\mu &=& \frac{\sqrt{2}v_t m_A^2}{v_d^2 + 4v_t^2}\,, \label{eq:lambda5}
\end{eqnarray}
\label{eq:quartics}
\end{subequations}

where, $s_\alpha(c_\alpha) = \sin \alpha(\cos \alpha)$. Among all these, the electroweak vev and the 
lightest CP-even Higgs mass $m_h = 125~\rm GeV$ are known. The other remaining non-standard
scalar masses and the mixing angle of CP-even scalars $\{m_H, m_A, m_{H^\pm}, m_{H^{\pm\pm}}, \alpha \}$ are our 
set of free independent parameters which also serve as the boundary conditions
for the RG evolution of the scalar quartic couplings for a fixed triplet vev $v_t$.


\section{Theoretical and Experimental constraints}\label{constaints}

To have an absolutely stable potential up to very high scale, the following
set of theoretical constraints must be satisfied at each scale of
RG running starting from the EW scale up to the high cut-off scale (GUT or Planck scale). 
In addition 
to this, one must also ensure the 
non-violation of the electroweak precision test data as well as the existing collider bounds 
on both singly and doubly charged scalars. The lightest neutral CP-even state being considered 
as the SM-like Higgs, the corresponding Higgs signal strengths should be also consistent with the 
current experimental limits from the 13 TeV LHC data in the model parameter region. 

\begin{itemize}

\item {\bf {Vacuum stability}} 

To have the scalar potential to be bounded from below in all direction in field space,
the following necessary and sufficient conditions
on the scalar quartic couplings has to be satisfied \cite{Arhrib:2011uy}\footnote{In a recent study~\cite{Bonilla:2015eha}, it has been shown that these
conditions can be further relaxed. In this work, we consider the restrictive case.}:
\begin{subequations}
\begin{eqnarray}
 \lambda  &\geq& 0\,, \label{eq:st1}\\
 \lambda _2+\lambda _3 &\geq& 0\,, \label{eq:st2}\\
 \lambda _2+\frac{\lambda _3}{2} &\geq& 0, \label{eq:st3}\\
\lambda _1+ \sqrt{\lambda (\lambda _2+\lambda _3)} &\geq& 0\,, \label{eq:st4}\\
 \lambda _1+ \sqrt{\lambda \left(\lambda _2+\frac{\lambda _3}{2}\right)} &\geq& 0, \label{eq:st5}\\
 \lambda _1+\lambda _4+\sqrt{\lambda (\lambda _2+\lambda _3)}&\geq& 0\,, \label{eq:st6}\\
  \lambda _1+\lambda _4+ \sqrt{\lambda \left(\lambda _2+\frac{\lambda _3}{2}\right)} &\geq& 0 \label{eq:st7}.
\end{eqnarray} 
\label{eq:lstab}
\end{subequations}
\item {\bf Perturbative unitarity} 

Furthermore, the tree-level unitarity of the 
scattering amplitude for all $2\to2$ processes demands the ${\cal S}$-matrix eigenvalues to be bounded from above as below \cite{Arhrib:2011uy}:
\begin{subequations}
\begin{eqnarray}
|\lambda _1+\lambda _4| &\leq & 16\pi\,, \label{eq:uni1} \\
|\lambda _1| &\leq& 16\pi\,, \label{eq:uni2}\\
|2\lambda _1+3\lambda _4| &\leq& 32\pi\,, \label{eq:uni3}\\
|\lambda | &\leq& 32\pi\,, \label{eq:uni4}\\
|\lambda _2| &\leq& 8\pi\,, \label{eq:uni5}\\
|\lambda _2+\lambda _3| &\leq& 8\pi\,, \label{eq:uni6}\\
|\lambda +4\lambda _2+8\lambda _3\pm\sqrt{(\lambda -4\lambda _2-8\lambda _3)^2+16\lambda^2_4}| &\leq& 64\pi \,, \label{eq:uni7}\\
|3\lambda +16\lambda _2+12\lambda _3\pm\sqrt{(3\lambda-16\lambda _2-12\lambda _3)^2+24(2\lambda _1+\lambda _4)^2}| &\leq& 64\pi\,, \label{eq:uni8}\\
|2\lambda _1-\lambda _4| &\leq& 16\pi\,,\label{eq:uni9}\\
|2\lambda _2-\lambda _3|&\leq& 16\pi. \label{eq:uni10}
\end{eqnarray}
\label{eq:luni}
\end{subequations}

\item {\bf Constraints from electroweak precision test} 

The new triplet Higgs bosons contribute to the electroweak precision 
observables, namely, the S,T, U parameters \cite{Lavoura:1993nq, Aoki:2012jj}. The strongest
bound comes from the T-parameter which imposes strict limit on the 
mass splitting between the doubly and singly charged scalars, 
$\Delta M \equiv \mid m_{H^{\pm \pm}} - m_{H^\pm}\mid $ which
should be $\lesssim 50$ GeV~\cite{Chun:2012jw,Baak:2014ora}
assuming a light SM Higgs boson of
mass $m_h = 125$ GeV and top quark mass $M_t = 173 $ GeV.
In our study, we do not explicitly analyze the T-parameter
constraints but rather impose this restrictive bound of 50 GeV between the two charged scalar masses.

\item {\bf Experimental bounds on scalar masses } \\
Several heavy scalars predicted by this scenario can be 
probed at both $e^+e^-$ and hadron colliders. The direct search on the 
singly charged scalar at the LEP II puts a limit on $m_{H^\pm}\geq 78 $ GeV 
\cite{Searches:2001ac}. For our all benchmark points, the charged scalar masses 
vary between $\sim 173~{\rm GeV}-180~{\rm GeV} $ and for this mass range, the 
$ t \to b H^{+}$ decay mode is kinematically suppressed, hence, 
there is no experimental limit on our chosen benchmark points 
from the charged Higgs search in $ t \to H^+ b$ followed by the decay
$H^+ \to c {\bar s}$ at the LHC~\cite{Aad:2012rjx,Chatrchyan:2012vca}.

At the LHC, the dominant production mode for a heavy neutral 
Higgs boson is the gluon fusion process: 
$g g \to \Phi $, where $\Phi \equiv H/A$. 
In Type II seesaw model, interactions between quarks and $SU(2)_L$ triplet
scalars happen via the doublet (SM like Higgs boson) 
and triplet mixing, which is proportional to $(v_t/v_d)$.
For our choice of triplet vev $(v_t = 3~{\rm GeV})$, the $\sigma (pp (gg) 
\to H)$ is proportional to $(v_t/v_d)^2 \sim {\cal O}(10^{-4})$. 
As a result of this large suppression the production cross-section of the 
heavy neutral scalar $(H)$, $\sigma ( pp (gg) \to H) $ at 13 TeV LHC is 
approximately 13 (fb)\footnote{We use the HIGLU code \cite{Spira:1995mt} to 
generate the $\sigma^{\rm NNLO QCD+EW} (pp \to H)$ and then rescale it with 
doublet-triplet mixing factor to obtain the $\sigma (pp \to H)$
in this scenario.} for the best possible benchmark point and this is 
well below the current $95\%$ CL bound on $\sigma (gg \to H) \times 
{\rm BR}(H \to ZZ)({\rm pb})-m_H$ plane by the ATLAS collaboration~\cite{TheATLAScollaboration:2016bvt}. One can draw similar 
conclusions for the case of heavy neutral pseudo scalar $(A)$ case also.
The second most dominant process for the heavy neutral Higgs production at the
LHC is through vector boson fusion (VBF). The ATLAS Collaboration searched the
signal of heavy neutral Higgs boson through the production via VBF, followed 
by its decay into pair of vector bosons $(W^+W^-)$ and they ~\cite{Aaboud:2017gsl} put a $95\%$ CL bound on
$\sigma ({\rm VBF} \to H) \times {\rm BR}(H \to W^+W^-)$ pb - $m_H$ ( GeV) plane.
We have checked that all our benchmark points are well within this experimental
limit. The collider bound on the doubly charged Higgs boson 
mass strongly depends on the triplet vev $v_t$. For $v_t < 10^{-4}$ GeV
(corresponds to large Yukawa couplings) and assuming degenerate scalars, 
the doubly charged Higgs boson decays to like sign dilepton (LSD) 
$\ell^\pm \ell^\pm $ with almost $100\%$ probability. 

From the direct search of the doubly charged Higgs 
boson at 13 TeV LHC run, the current lower bound at $95\%$ CL on 
its mass is $m_{H^{\pm\pm}}  > 700-800 $ GeV~\cite{ATLAS:2017iqw} depending upon the 
final state lepton flavour. On the other hand, for $v_t > 10^{-4}$ GeV
( corresponds to small Yukawa couplings), in our case $v_t = 3 $ GeV, the 
branching ratio into LSD decreases substantially and there are other 
several competing decay modes of the doubly charged Higgs boson open up,
like (i) pair of heavy charged gauge bosons $(W^\pm W^\pm)$, (ii) $W^\pm H^\pm$
and (iii)$ H^\pm H^\pm$, if kinematically accessible. Due to the cascade nature
of the final state, the collider bound on the doubly charged Higgs boson in
this case is rather weak and we take it to about 100 GeV \cite{Melfo:2011nx}.

\item {\bf Constraints from Higgs signal strength} 

The lightest CP-even state resembles the SM Higgs boson of mass 125 GeV, the decay widths of
which should be in concurrence with the currently available Higgs 
data from the LHC. Following the aforementioned argument, the suppressed
mixing between the doublet and triplet scalars lead to negligible contribution to the dominant
production(gluon fusion) channel and to the tree-level decay widths of the Higgs. However, the total
width will mainly be modified by the loop-induced decay modes $h\to \gamma \gamma$  where the 
non-standard singly and doubly charged scalar may contribute significantly. Some detailed studies have
already been performed in this respect~\cite{Chun:2012jw,Aoki:2012jj,Arbabifar:2012bd,Kanemura:2012rs,Akeroyd:2012ms,Dev:2013ff,Das:2016bir}. Here we will refrain
ourselves from giving an elaborate description except a comment on the parameter space and
for the collider study, our choice of benchmark points remain within the $2\sigma$ limit
of the current experimental bound of the Higgs to diphoton signal strength~\cite{ATLAS:2016nke}.
\end{itemize}
\section{High-scale stability}\label{sec:running}

In this section, we set out to describe the high-scale nature of the model scenario. 
To find out the high-scale valid region of the parameter space, we analyze the 
one-loop RG running of all the scalar 
quartic coupling together with
the gauge and Yukawa (mainly top) coupling from the EW to some high scale\footnote{The contribution to low energy effective potential
	for the SM Higgs doublet is negligible~\cite{Dev:2013ff} and thus the threshold effect has been safely ignored.}.
The corresponding beta functions are given in the Appendix. 
We choose two distinct values
of the triplet vev, 1 GeV and 3 GeV respectively. In this regard, we should mention that
although, in the first place, the triplet vev can acquire any value between eV to few GeV range, we only consider the values at GeV range in regard to our collider studies, described in later sections. 
Also one more important point is that such large $v_t$ leads to 
small Yukawa couplings for the neutrinos and therefore,
we can safely ignore the effect of the running of neutrino Yukawa coupling 
in the RGE~\cite{Chao:2006ye,Dev:2013ff}. 
In our RGE analysis we scan the parameter space in the following range:
\begin{eqnarray}
m_H(m_A) &\in& \{m_h, 2000 \} \rm GeV \,, \qquad  \sin \alpha \in \{-0.1, 0.1 \} \nonumber \\ 
 m_{H^\pm}(m_{H^{\pm\pm}}) &\in& \{100, 2000\} \rm GeV \,, 
\label{eq:rangescan}
\end{eqnarray}
while we fix $m_h=125 ~\rm GeV$, the EW vev $(v)=246 ~\rm GeV$. The top quark pole mass 
is set at $M_t=173 ~\rm GeV$ for which the running top mass is $m_t(M_t)=164 ~\rm GeV$. 
All the scalar quartic couplings are then derived using Eq.~(\ref{eq:quartics})
to set their boundary conditions at  $M_t$
and we run the full one-loop RGE (see Appendix)
from $M_t$ to some high cut-off scale. To ensure the 
stability and perturbative unitarity of the potential, we check that the conditions
stated above in Eqs.~(\ref{eq:lstab}) and (\ref{eq:luni}) do not violate at any 
scale during the running from EW to some high scale (GUT or Planck) up to which the validity of the
theory is tested.
\begin{figure}[ht!]
	\centering
		\includegraphics[width=80mm,height=60mm]{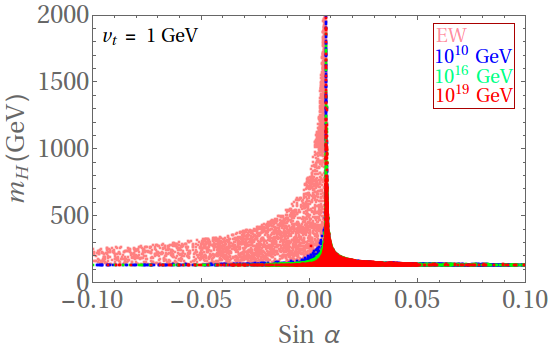} 
		\includegraphics[width=80mm,height=60mm]{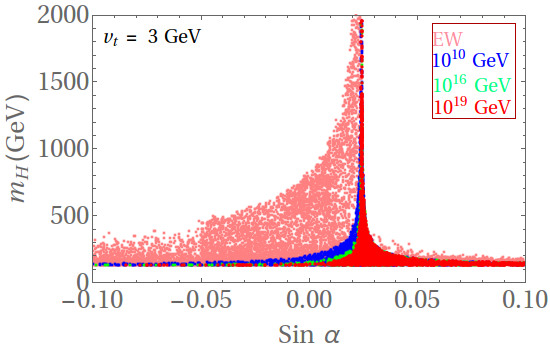} \\
		\includegraphics[width=80mm,height=60mm]{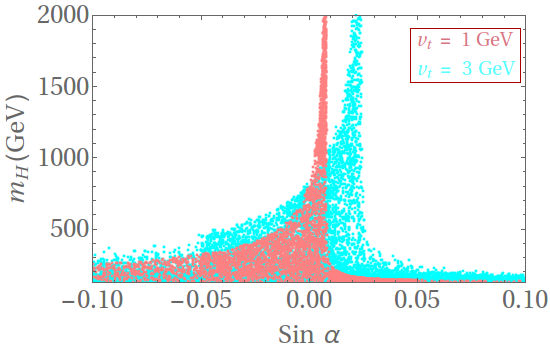} 
		\includegraphics[width=80mm,height=60mm]{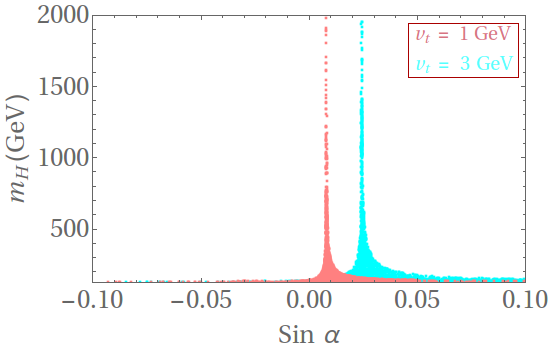}
	\caption{\label{fig:stab} \textit{(Upper panel) The valid parameter space in the $\sin\alpha-m_H$ plane for
			$v_t=1~\rm GeV$ (left) and $v_t=3~\rm GeV$ (right) for different values of cut-off scale. Different colors indicate the different high
			cut-off scale while the background (pink) shaded region is only allowed at the EW scale.
			(Lower panel) The explicit distinction between the allowed parameter space by the two different $v_t$ valid
			only at the EW scale(left) and all the way up to the Planck scale(right).}}
\end{figure}

Henceforth, whenever we mention a high-scale valid region
we would mean the parameter space allowed by both the high-scale stability and the perturbativity constraints.
This follows from the fact that
the scalar quartic couplings do not feel any negative pull from fermionic loops during its running as the
triplet does not couple with quarks and the lepton Yukawas are too small to consider. 
Therefore, the strongest constraint will come from the demand of high-scale perturbative unitarity.
Although, one can always think of a model that does not respect perturbative unitarity at high scale
which can call for some UV completion of the model, we restrict us from discussing this possibility
in this work. 

Now, as mentioned before, we have also considered a restrictive bound of 50 GeV between 
the singly and doubly charged scalar mass splittings, not to violate the T-parameter constraints. 
But, the requirement of absolute stability and unitarity actually
entail a relation among the scalar mass parameters and not all of them 
remain independent. This is true even at the EW scale~\cite{Das:2016bir},
however the extra demand of high-scale validity puts more stringent bound on
the limits. 
For the sake of completeness, we would like to briefly explain
the inherent nature of the parameter space due to the simultaneous stability
and unitarity constraints. The conditions given in Eqs.(\ref{eq:st1} and \ref{eq:uni6})
when translated to mass terms using Eq.(\ref{eq:quartics}), 
yield the CP-odd neutral scalar mass $(m_A)$ as a function 
of the two other CP-even scalar masses $(m_h,m_H)$ and their mixing angle $\alpha$ and can be approximated as,

\begin{eqnarray}
m_A^2 \simeq (m_H^2 \cos^2\alpha + m_h^2 \sin^2\alpha)\, 
\label{eq:mA}
\end{eqnarray}
for $v_t << v_d$.
In the similar fashion, the doubly charged scalar mass can also be determined 
once we set the unitarity condition of Eq.(\ref{eq:uni10}) in addition to the above
which turns out to be in the close proximity of $(2m_{H^\pm}^2 - m_A^2)$. 
We did not use this mass relations for the $m_A^2$ and $m_{H^{\pm\pm}}^2$ 
to scan the parameter range instead we independently scan all the masses in the range
given in Eq.~(\ref{eq:rangescan}).
But, at the end, we see that the mass of $m_A$ and $m_{H^{\pm\pm}}$ bear such relation to 
maintain the stability and unitarity constraints. 
In the first row of Fig.{\ref{fig:stab}}, we show the allowed region for the
EW scale and the three distinct cut-off scales
namely, $10^{10} \,, 10^{16} ~\&~ 10^{19}$ GeV respectively in $(\sin \alpha$-$m_H)$ plane, 
where $m_H$ denotes the mass of heavier CP-even Higgs. 
It is interesting to note that significant amount of allowed parameter space at EW scale shrinks once we impose the 
stability of the vacuum all the way up to the Planck scale.

It is also worth noticing that
the large value of the exotic scalar masses happens only for a small nonzero value
of $\sin\alpha$ when high scale stability is demanded and the
absolute range of $\sin\alpha$ shifts towards more positive value if the triplet vev is
increased from 1 GeV to 3 GeV. This is again a consequence of the unitarity bound which
can be understood by looking at Eq.~(\ref{eq:uni1}). 
The relation when translated in terms of the physical scalar masses using Eq.~(\ref{eq:quartics}),
 turns out to be
\begin{eqnarray}
 \frac{2 m_{A}^2}{v_d^2 + 4 v_t^2} + \frac{\sin 2\alpha}{2 v_d v_t}(m_h^2 - m_H^2) < 16 \pi \,. 
\end{eqnarray}
Now, for $v_t<<v_d$ and using Eq.~(\ref{eq:mA}), the limit is trivially satisfied for $\sin\alpha \simeq 2v_t/v_d$ 
reaching the decoupling limit of large $m_H (>> m_h)$. This also means that the in the SM-like limit for large $m_H$, the mixing angle tends to zero for $v_t \to 0$.
Therefore, with an increase in $v_t$ will essentially shift 
the decoupling region for the non-standard scalars to larger mixing angle. This feature is
reflected in Fig.~\ref{fig:stab} where the peak of the allowed region is shifted for $v_t= 1$ to 3 GeV
to more positive $\sin \alpha$.
In the second row of Fig.~\ref{fig:stab}, we explicitly show the shift in peak for the two different values of the $v_t$ 
in the region allowed only at the EW scale and all the way up to the Planck scale respectively.
\begin{figure}[t!]
	\centering
		\includegraphics[width=80mm,height=60mm]{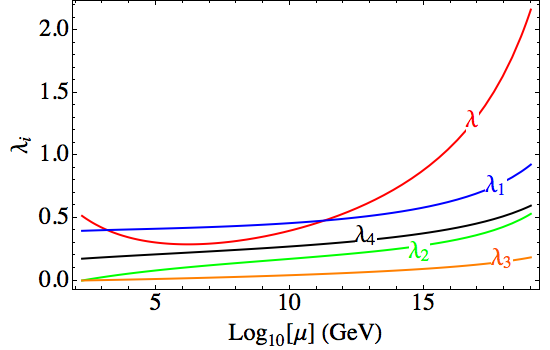} 
		\includegraphics[width=80mm,height=60mm]{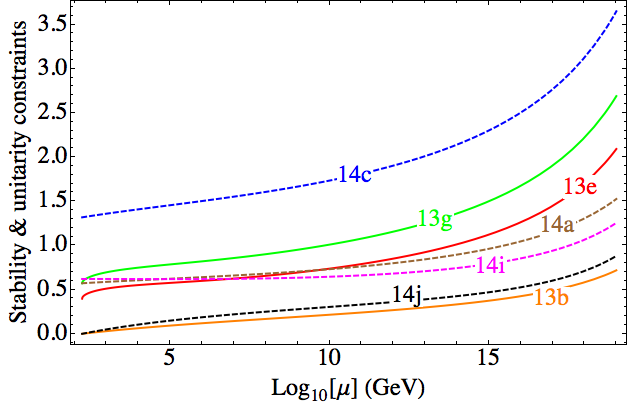}
	\caption{\label{fig:runnning} \textit{(Left panel) The running of the five scalar quartic couplings up to the
			Planck scale for the benchmark point BP1 of positive scenario as given in Table~\ref{table:bp}.
			(Right panel) The running of some of the stability and unitarity constraints indicated by the
			corresponding equation numbers for the same benchmark point.}}
\end{figure}

We  also show the running of individual scalar quartic coupling and some of the stability 
and unitarity conditions in Fig.~\ref{fig:runnning} valid up to the Planck scale 
for a benchmark point (BP1 of positive scenario) given in Table~\ref{table:bp}.

In the next part of this paper, we shall explore the parameter space
available at the Planck scale allowed region in the current LHC run
and look at the prospect of some possible collider signals for the non-standard
scalar particles.  
Accordingly, it is to be noted from Fig.~\ref{fig:stab} (upper two panels) 
that for large mixing angle $(\sin\alpha \sim {\cal O}({0.1}))$, a 
Planck scale valid region only allows the heavier neutral scalar mass $(m_H)$ 
close to the lighter SM-like higgs mass $m_h$, namely the degenerate scenario. 
Moreover, the other triplet scalar masses $(m_A,m_{H^\pm},m_{H^{\pm \pm}})$ are also pushed in the same mass range
due to the unitarity and 
T-parameter constraints. However, for our collider study we only consider the 
non-degenerate scenario\footnote{The phenomenology of the degenerate scenario 
has been studied in Ref.~\cite{Arbabifar:2012bd}.} and therefore a triplet scalar masses around a few hundred GeV (200-300 GeV)
instantly restrict us to a small range of mixing angle 
$(0.01 < \sin\alpha < 0.05)$ for $v_t = 3~ \rm GeV$ as can be seen 
from Fig~\ref{fig:stab}.

\begin{figure}[ht!]
	\centering
		\includegraphics[width=80mm,height=62mm]{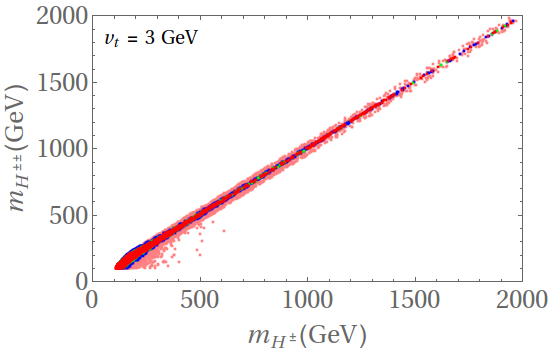} 
		\includegraphics[width=80mm,height=60mm]{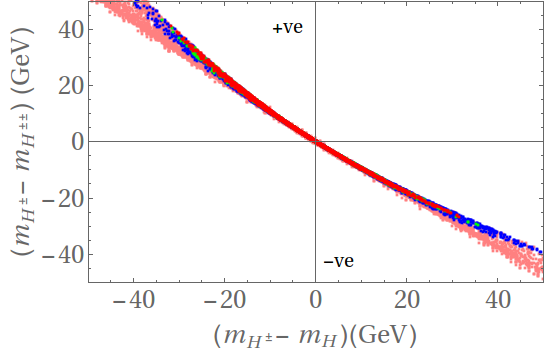}
	\caption{\label{fig:scenario} \textit{(Left panel) The allowed parameter space in the $m_{H^\pm}-m_{H^{\pm\pm}}$ 
			plane for triplet vev $v_t=3~\rm GeV$. The different colors follow the same convention as in
			Fig.~\ref{fig:stab}. (Right panel) The corresponding allowed parameter space to show the relation between the mass splittings 
			of the singly charged Higgs to the neutral Higgs $(m_{H^\pm}-m_H)$ and singly charged Higgs to the doubly charged Higgs $(m_{H^\pm}-m_{H^{\pm\pm}})$.
			The upper left square corresponds to the valid region for our positive scenario 
			while the lower right corner denotes the same but for our negative scenario. }}
\end{figure}
In this work, we aim to study the possible signature of charged scalar sector of the model, more
specifically, we investigate the signal of the associated production of the singly 
and doubly charged scalars for some specific signal processes. Before going into 
the detail, we show the allowed parameter space in $(m_{H^\pm}-m_{H^{\pm\pm}})$ plane
in the left panel of Fig.~\ref{fig:scenario} for $v_t=3 ~\rm GeV$. Here, one should
recall that the doubly charged mass in the high-scale stable parameter space
is related to the singly charged scalar and neutral scalar mass squares as:  
\begin{eqnarray}
m^2_{H^{\pm\pm}}-m^2_{H^{\pm}}\approx m^2_{H^{\pm}}-m^2_{A}(m^2_{H})\,.
\label{eq:massdiff}
\end{eqnarray} 
This yields two different scenarios depending on the mass hierarchy between the triplet scalars:
\begin{enumerate} 
\item Positive scenario $(\lambda_4 > 0)$ :: $m_{H^{\pm\pm}}< m_{H^{\pm}} < m_{A}/m_{H}$.
\item Negative scenario $(\lambda_4 < 0)$ :: $m_{H^{\pm\pm}}> m_{H^{\pm}} > m_{A}/m_{H}$.
\end{enumerate} 

In fact the names are self-explanatory since the mass differences in
Eq.~(\ref{eq:massdiff}) equate to $ -\frac{1}{4}\lambda_4 v_d^2$ which is obtained from
Eq.~(\ref{eq:lambda4}). Therefore, positive scenario stands for a positive $\lambda_4$
while the negative scenario delivers a negative $\lambda_4$. In the right panel
of Fig.~\ref{fig:scenario}, we show the high-scale allowed parameter space for these
two distinct mass scenarios. We shall perform the collider analysis for both
mass scenarios. 
Now, at this point, we would like to comment that we have explicitly checked that 
the allowed region up to the Planck scale is consistent with the Higgs to diphoton signal strength at $2\sigma $ level.

\section{Collider Analysis}
\label{sec:collider}

\begin{figure}[htbp!]
\centering
\includegraphics[width=80mm,height=62mm]{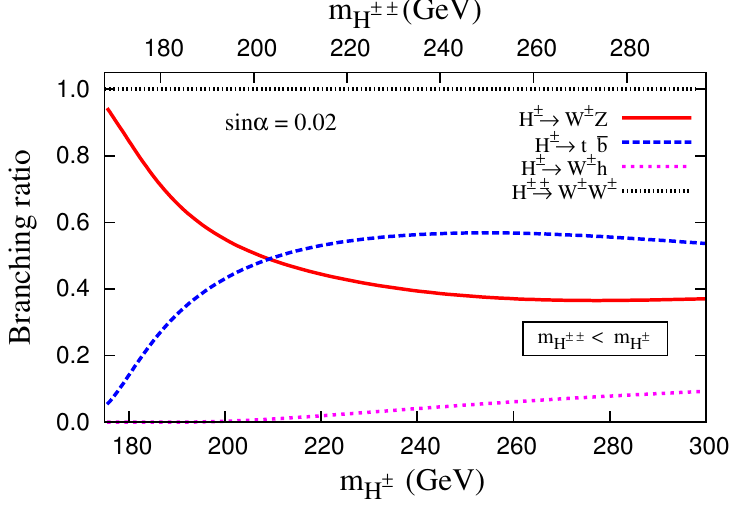}
\caption{\textit{Branching ratio of different two body decay modes for the doubly charged and singly charged 
scalars for $\sin\alpha  = 0.02$ and $v_t =3~\rm GeV$ for the positive scenario.}}
\label{fig:br} 
\end{figure}

Following our discussion in the previous section, hereby, we address some
predictive collider signal in the charged
scalar sector at the current LHC run with $\sqrt{s}= 13~\rm TeV$,
in compliance with the parameter space available when the cutoff scale is
set at the Planck scale.
We mainly pursue the associated production of the doubly and singly charged 
scalar as our dominant production channel\footnote{The neutral scalars 
with similar mass as the singly and doubly charged scalars have comparable production 
cross-section ($H ~\rm A$) with the aforementioned pair and associated production processes. This
has been studied in detail in ~\cite{Han:2015sca}.}. To understand the ensuing
final states at the LHC, it is thus instructive to check the decay
patterns of the charged scalars in the chosen parameter space. 
We have mentioned earlier that at large $v_t\sim 3~\rm GeV$, the leptonic decay 
modes of the triplets become extremely suppressed and only
decay to gauge bosons are allowed.
At such value, doubly charged Higgs 
 decay to $W^\pm W^\pm$ with almost 100\% branching ratio while the
 singly charged scalar can have mainly three types of 
 decay modes $(W^\pm Z, t\bar b, W^\pm h)$ depending on the phase space available due to its mass. In Fig.~\ref{fig:br},
 we show the variation of branching ratio for both the charged scalars as a function of $m_{H^{\pm}} $ and $m_{H^{\pm\pm}}$
 for the mixing angle $\sin\alpha=0.02$. Here only the
 positive mass scenario has been depicted, the decay patterns remain unchanged for the negative scenario.
 We only show the mass ranges above the $W^\pm Z$ onshell mass threshold below which
 only three body decay modes are available for both singly and doubly charged scalars.
 Depending on the mass difference between the charged scalars either off-shell gauge boson ($W^\pm$ or $Z$)
 or off-shell charged scalar $H^{\pm \pm}$($H^{\pm}$) will be produced for 
 positive(negative) scenarios. Therefore, we only show the mass ranges ( $>$ 170 GeV ) that allow two body decay modes of the
charged scalars.
 
For $ (170-200) ~\rm GeV$ mass range, the singly charged scalar dominantly decays into
$W^\pm Z$ mode. On the other hand, the doubly charged scalar decays to $W^\pm W^\pm$ with 100\% branching ratio
due to the choice of the triplet vev ($v_t$).
The presence of the multi gauge bosons ($W^{\pm},Z$) in the final state motivate us to consider the following two signal topologies
\footnote{There are four additional signal topologies. The one with $4\ell^\pm + 2j + \mET$ 
	has been studied in a recent article \cite{Mitra:2016wpr} and will not be mentioned
	in this work.The other three possible final states $5\ell^\pm + \mET$, $3\ell^\pm + 4j + \mET \,$ and 
	$(\ell^+\ell^-) + 4j + \mET$ should also be present but with much less signal significance and is not comparable
	to the above cases. We will briefly mention them again in subsequent places.}:
\begin{eqnarray}	
&(i)&    3\ell^\pm + \mET \,; \nonumber  \\
&(ii)&	 (\ell^+\ell^+) + 4j + \mET \,, \nonumber 
\label{final_states}
\end{eqnarray}
where, $\ell = e, \mu $ and $j$ corresponds to jets (included non-tagged
$b$-jets). We also include the contributions from the {\it charge conjugated} process. 
At this point, it is worth mentioning that the experimental 
searches done till date has only scrutinized the possibility of
having multilepton final states from the primary decay of 
the doubly charged scalar probing only the lower triplet
vev ( $v_t \leq 10^{-4}~\rm GeV $ ) scenario. However, 
those search strategies can not be applicable
for the signal processes (i) and (ii) due to very different kinematics of the final 
state leptons which are the end product of the cascade decay of singly and 
doubly charged scalars. Through our detailed analysis we will show that our principal search mode for this
scenario relies on the final state (ii). Here it should also be mentioned that while the final state (i) can only emerge 
from the associated ($pp\rightarrow H^{\pm\pm} H^{\mp}$) production process, 
the other final state (ii) receives contributions from both the associated ($pp\rightarrow H^{\pm\pm} H^{\mp}$)
and the pair-production ($pp\rightarrow H^{++} H^{--}$) processes.
\begin{figure}[htbp!]
\centering
\includegraphics[width=80mm,height=62mm]{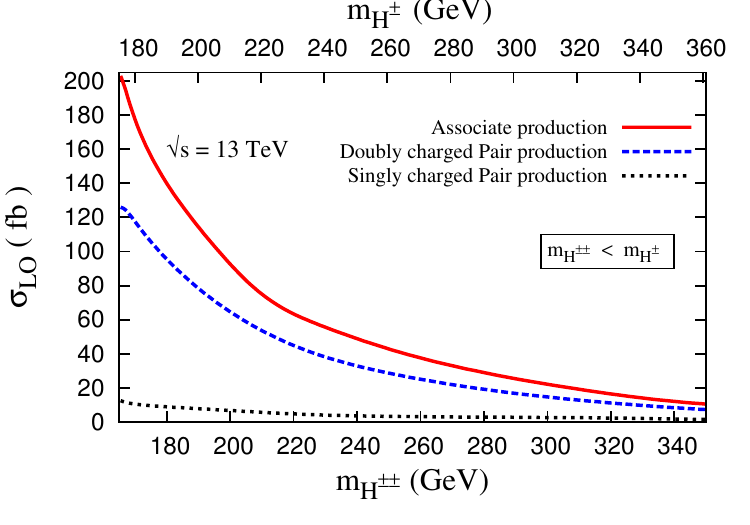}  
\includegraphics[width=80mm,height=62mm]{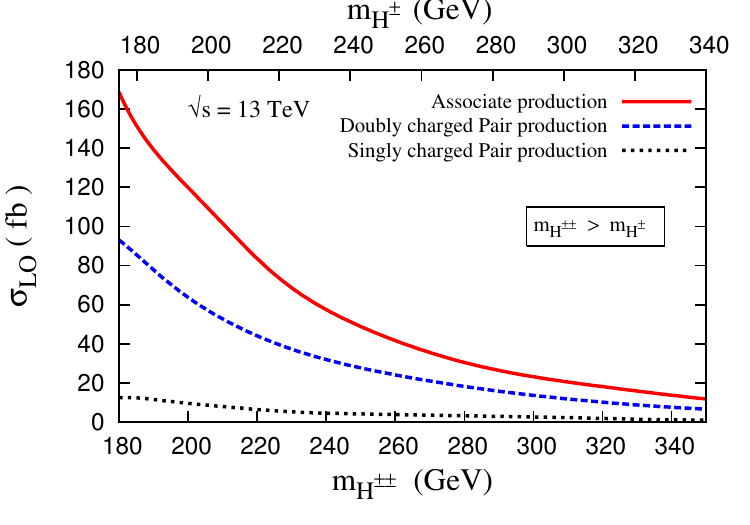}
\caption{\textit{ Left (Right) panel shows the variation of $\sigma_{\rm LO}(pp\rightarrow H^{\pm\pm} H^{\mp}) ~\rm (fb)$ (solid red curve),
$\sigma_{\rm LO}(pp\rightarrow H^{++} H^{--}) ~\rm (fb)$ (blue dashed curve) and $\sigma_{\rm LO}(pp\rightarrow H^{+} H^{-}) ~\rm (fb)$
(black dotted curve) with respect to charged Higgs masses at the LHC at $\sqrt{s}= \rm 13~TeV$ for positive (negative) scenario.}}
\label{fig:cros} 
 \end{figure}

In Fig.\;\ref{fig:cros}  we show the $\sigma_{\rm LO}(pp\rightarrow H^{\pm\pm} H^{\mp})$, 
$\sigma_{\rm LO}(pp\rightarrow H^{++} H^{--})$ and $\sigma_{\rm LO}(pp\rightarrow H^{+} H^{-})
$ cross sections at the $\sqrt{s} = 13 ~\rm TeV$ LHC (where LO stands
for leading order). As it is very evident from these figures that the pair
production rate of singly charged scalars is order of magnitude smaller than 
that of $H^{++}H^{--}$ and $H^{\pm\pm}H^{\mp}$ production 
cross-sections. Hence, for our collider study we only consider 
$H^{++}H^{--}$ and $H^{\pm\pm}H^{\mp}$ production processes.
\begin{table}[ht!]
	\centering
\resizebox{10cm}{!}{
	\begin{tabular}{|c|c|c|c|c|c|}
		\hline
		Mass Scenario & $\sin\alpha$ & $m_{H^{\pm\pm}}$ & $m_{H^{\pm}}$ & $m_H=m_A$  & {$\mu_{\gamma\gamma}$}  \\  
		 &  & $(\rm GeV)$ & $(\rm GeV)$ & $(\rm GeV)$ &{} \\  \cline{1-1}
        Positive & & & & & \\ \hline
        BP1 & 0.0220 & 165.48 & 173.25 & 180.70 & 0.79  \\ \hline
                BP2 & 0.0280 & 175.99 & 177.47 & 178.93 & 0.82  \\ \hline\hline 
        	Negative & & & & &  \\ \cline{1-1}
      		BP1 & 0.0277 & 179.60 & 176.30 & 173.01 & 0.79 \\ \hline
            BP2 & 0.0300 & 184.17 & 180.11 & 175.95 & 0.81 \\ \hline
        	
		\end{tabular}}
	\caption{\it Benchmark points valid by the high-scale stability constraints
	up to the Planck scale and their corresponding Higgs to diphoton signal strength ($\mu_{\gamma\gamma}$) for both the positive 
	and the negative scenario. The current experimental value of $\mu_{\gamma\gamma}$ is $0.85^{+0.22}_{-0.20}$~\rm \cite{ATLAS:2016nke}.}
	\label{table:bp}
\end{table}

We now choose few representative benchmark points as shown in 
Table ~\ref{table:bp} from the region which are allowed by the stability 
condition all the way up to the Planck scale. We show benchmark 
points for both the mass hierarchies.

Along with the masses of the triplet scalar
and the neutral mixing angle, we present the Higgs to diphoton signal strength ($\mu_{\gamma\gamma}$)
for each benchmark points which is allowed by the $2\sigma$ limit of the current Higgs data 
$(0.85^{+0.22}_{-0.20})$~\cite{ATLAS:2016nke}.

For our analysis, both the signal and SM backgrounds events are generated at the Leading Order (LO) 
parton level in {\tt Madgraph5(v2.3.3)}~\cite{Alwall:2014hca} using
the {\tt NNPDF3.0} parton distributions~\cite{Ball:2014uwa}. The model has been implemented in {\tt FeynRules}~\cite{Alloul:2013bka}
which gives the UFO model files required in madgraph. The parton showering and hadronisation is done using the built-in {\tt Pythia}
\cite{Sjostrand:2006za} in the madgraph. The showered events are then passed through {\tt Delphes}(v3)~\cite{deFavereau:2013fsa} 
for the detector simulation where the jets are constructed 
using the anti-$K_{T}$ jet algorithm. For the background processes with
hard jets, proper MLM matching scheme \cite{Hoche:2006ph} has been chosen.
The cut-based analyses are done using the {\tt MadAnalysis5}~\cite{Conte:2012fm}.

Several SM processes contribute as backgrounds to the aforementioned two 
final states. We consider the following SM processes in our analysis: 
$t\bar{t}$+jets (up to 3), single top with three hard jets, 
$V+{\rm jets~(up~to~3~jets)}, V \equiv W^\pm, Z$, $VV$+ 3 jets,
$t\bar{t}+(W^\pm/Z/h)$, and  $VVV$. 
As we will see, the backgrounds from  
top quark production can be handled using the $b$-veto while the 
$W^\pm/Z + ~\rm jets$ and $W^+W^-+~\rm jets$ processes with large production cross-section
can be suppressed with the three-lepton or same-sign dilepton
selection criteria. Finally, the irreducible backgrounds left are
the $W^\pm Z +~\rm jets$ and $t\bar t+(W^\pm /Z/h)$ with small effective cross-section.
It is worthwhile to mention here that all the signal production channels are of purely electroweak type, 
while some of the SM background processes are either pure QCD or
QCD+EW in nature with huge cross section. However, as we will show, a suitable 
choice of selection cuts can improve the signal significance appreciably.

In our signal and background events, we select jets and leptons using the 
following basic kinematical acceptance cuts : 
\begin{subequations}
\begin{eqnarray}
\Delta R_{j j} > 0.6,~~~~\Delta R_{\ell \ell} > 0.4,~~~~\Delta R_{j \ell} > 0.7\,, \\
\Delta R_{bj} > 0.7,~~~~\Delta R_{b \ell} > 0.2,\\ 
p_{T_{\rm min}}^j > 20 ~{\rm GeV},~~~~|\eta_j| < 5,\\
p_{T_{\rm min}}^\ell > 10 ~{\rm GeV},~~~~|\eta_\ell| < 2.5,
\end{eqnarray}\label{basic_cuts}
\end{subequations}

where $\Delta R_{xy} = \sqrt{\Delta\phi_{xy}^2 + \Delta\eta_{xy}^2}$ ($x,y\equiv \ell, j, b$) and all other symbols have their usual meaning.

Our signal processes do not include any $b$-jets and therefore a veto on the 
$b$-tagged jets will reduce SM backgrounds with $b$-jets. Thus for the background
process, a jet has been tagged as a $b$-jet 
abiding by the efficiency as proposed by the ATLAS collaboration~\cite{ATLAS:2014pla}:
\begin{eqnarray}
\epsilon_b = \begin{cases}
0 & p_T^b \leq 30 ~\rm GeV \\
0.6 & 30 ~{\rm GeV} < p_T^b < 50 ~{\rm GeV} \\
0.75 & 50 ~{\rm GeV} < p_T^b < 400 ~{\rm GeV} \\
0.5 & p_T^b > 400 ~\rm GeV~.
\end{cases} 
\label{b-tag}
\end{eqnarray}

Also a mistagging probability of 10\% (1\%) for charm-jets (light-quark and gluon jets)
has been included. 
For the isolation of the leptons, we follow the criteria defined in Ref.~\cite{ATLAS:2016rin}
where the electrons are isolated with the {\tt Tight} criterion defined in Ref.~\cite{ATLAS:2016iqc} and the muons 
are isolated using the {\tt Medium} criterion defined in Ref.~\cite{Aad:2016jkr}


\subsection{Cut-based Analysis}
\label{subsec:cutanal}
In this section, we first examine the aforementioned signal topologies 
based on different kinematic variables. Then, considering the optimal
prospect, we propose some selection cuts to extract the signal from the
background. Finally, we shed light on the prospect of discovering such
signals at the LHC.

\subsubsection{ $3\ell^\pm + \mET \,$}

This final state consisting of trilepton plus missing transverse energy 
originates from the secondary decay of the gauge bosons 
produced from the decay of the charged scalars. For a 
particular charge assignment, the final state is developed as
\begin{eqnarray}
p p \to H^{++} H^- \to (W^+ W^+) + (W^-Z) \to (\ell^+ \ell^+) + \ell^- + \mET \,.
\end{eqnarray}
Even though this signal also encases a pair of same-sign lepton,
the presence of the third lepton hinders the process of selection.
Before detailing the selection cuts, let us first 
discuss the distributions of some of the relevant kinematic variables 
used in the signal selection procedure.
All the distributions are normalized to respective cross section of the
process and for the background distribution, we only show the irreducible 
background processes. One more important point to note
that we do not enforce the leptonic decay mode for the SM $W^\pm$ and $Z$ 
rather we consider the all inclusive decay channels for both the signal
and background event generation. This is true for all the subsequent analyses. 

\begin{figure}[htbp!]
	\centering
		\includegraphics[width=8cm,height=6cm, angle=0]{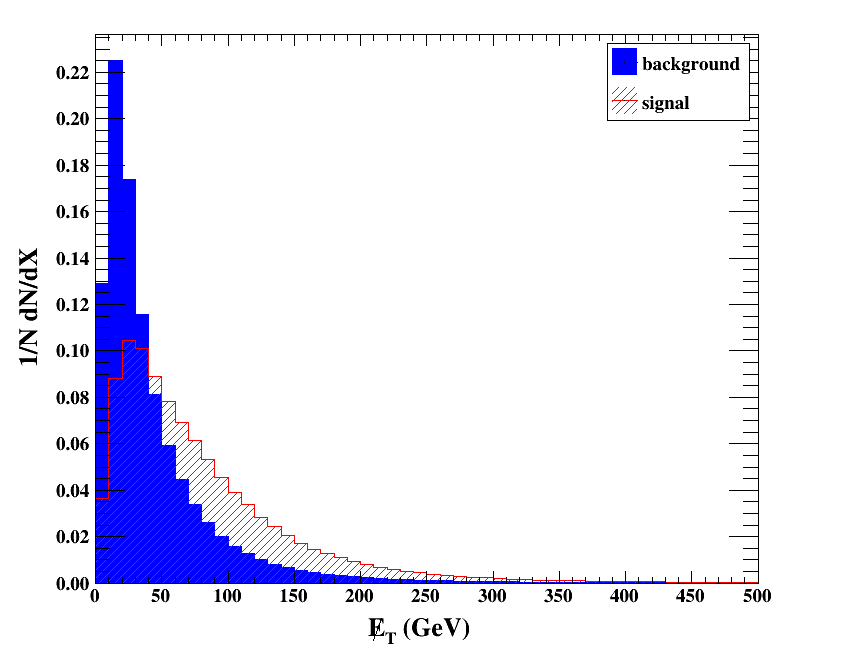}
		\includegraphics[width=8cm,height=6cm, angle=0]{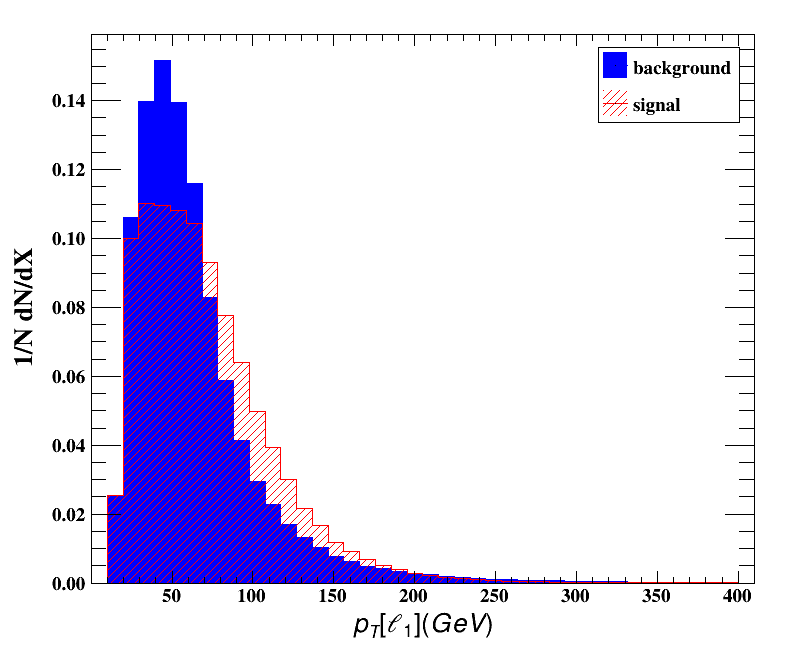}  
	\caption{\textit{ Normalized distribution of the (left panel) Missing transverse energy $(\mET)$  and
			(right panel) the transverse momentum of the hardest lepton $(p_T(\ell_1))$
			after the basic kinematical acceptance cuts for the benchmark BP1 of positive scenario.}}
	\label{fig:met} 
\end{figure}

We start with the missing transverse energy ($\mET$) 
distribution, depicted in the left panel of Fig.\;\ref{fig:met} for 
the benchmark point (BP1) of
the positive scenario with highest production cross-section. 
Since the decay pattern for the negative
scenario does not change, therefore we do not show the same for that scenario.
To explicate the nature of the distribution, we remind that
the $\mET$ ~for both the signal and the backgrounds originates 
solely from neutrinos (from $Z$ and $W^\pm$ decays). 
As a consequence, the overall nature of the histogram
for the signal and background events looks similar
except a slight deviation at the peak.
The background events (blue dark shaded area) 
peak at lower $\mET $ $(\lesssim 30 $ GeV) compared to our signal events. 
This is just an imprint of the slight boost in the gauge boson 
transverse momentum due to the heavier 
parent particles ( $H^{\pm \pm}$ and $H^\pm$). Hence, a requirement of 
$\mET > 30 $ GeV may be effective for a good signal to background ratio.

\begin{figure}[htbp!]
		\centering
		\includegraphics[width=8cm,height=6cm, angle=0]{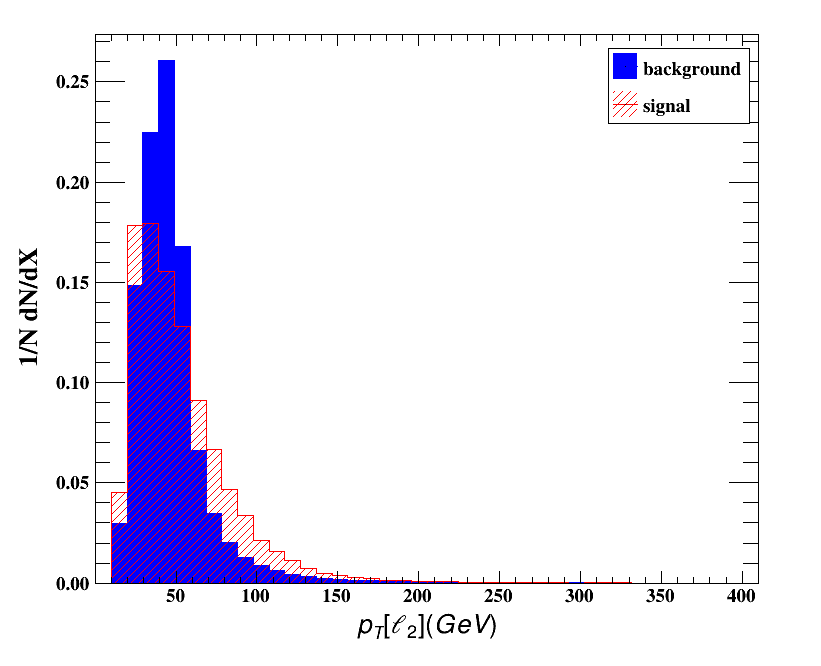} 
		\includegraphics[width=8cm,height=6cm, angle=0]{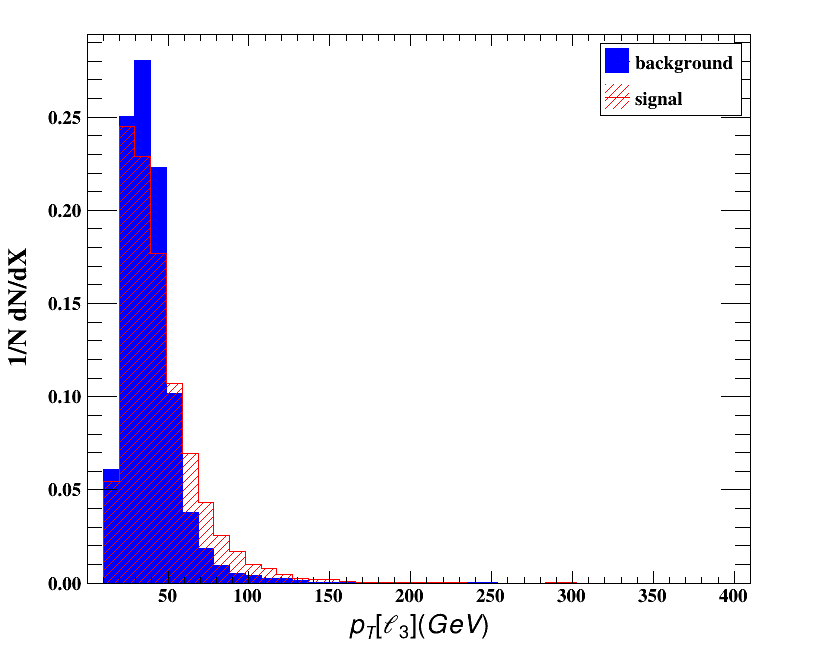}
		\caption{\textit{Transverse momentum $(p_T)$ distribution (normalized) of the two sub-leading leptons for the benchmark BP1 of positive scenario.}}
		\label{fig:3lep} 
	\end{figure}

In the right panel of Fig.~\ref{fig:met}, we show the transverse momentum ($p_T$) distributions 
of the hardest ($p_T$ ordered) lepton for the same benchmark while 
in Fig.~\ref{fig:3lep}, we draw the transverse momentum distribution
for the other two sub-leading leptons.
Again, for both the signal and background events, these three leptons
come from the SM gauge boson decay except with a little smearing effect
for the signal distribution due to its origin from heavier particles.
Therefore, it is quite difficult to put a selection cut on these
three leptons to disentangle the signal and background and we
can only put the basic acceptance cut to the three leptons
to assure a trilepton signature and veto on any additional fourth lepton.
However, one interesting fact is that the two opposite sign
leptons emanate from two different gauge bosons. This suggests
that an invariant mass cut on the same flavor opposite-sign lepton 
around the $Z$ boson mass shall surely reduce a large number of
background events from $W^\pm/Z+\rm jets$, $ZZ+\rm jets$ and $W^\pm Z+\rm jets$. 
In addition to this, since the signal is free from any additional jet,
a jet veto instantly cut down the giant background processes.

\begin{figure}[ht!]
		\centering
		\includegraphics[width=8cm,height=6cm, angle=0]{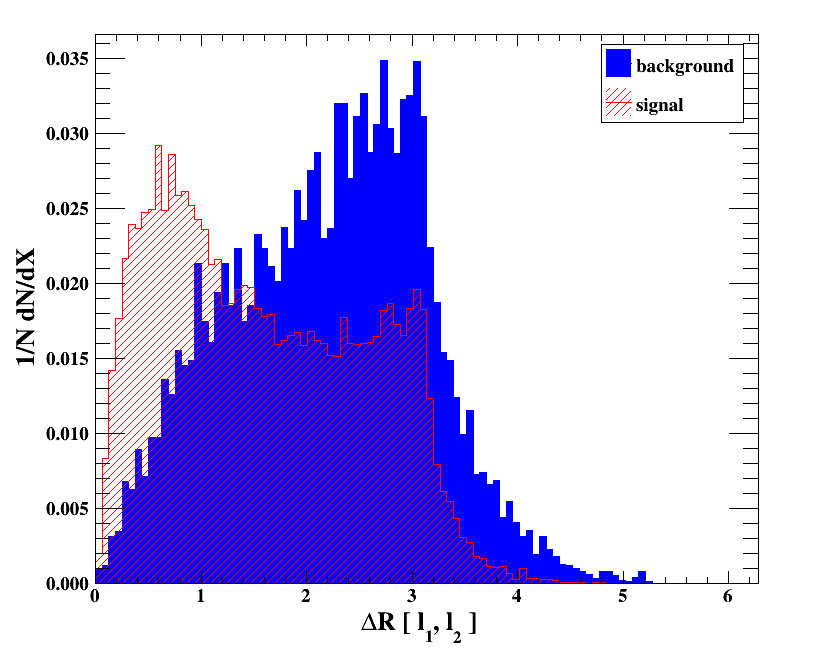} 
		\caption{\label{fig:dr} \textit{The $\Delta R(\ell_1^\pm \ell_2^\pm)$ distribution (normalized) between the two same-sign 
				leptons for the positive scenario benchmark BP1.
		}}
	\end{figure}

As already mentioned, the important feature of both of our signal processes
	is the presence of a pair of same-sign lepton, a claim of which can largely suppress the SM background.
	In this endeavor we define another kinematic variable, the angular separation $\Delta R(\ell^\pm\ell^\pm)$
	between the two same-sign leptons.
	The distribution is featured in Fig.\;\ref{fig:dr}. For our signal events, both these same-sign leptons come from
	the same-sign $W^\pm$ bosons which are produced from the decay of 
	single heavy doubly charged scalar $H^{\pm \pm}$. The same-sign 
	charged leptons tend to appear at small opening angle due to the spin 
	correlation between the parent $W^\pm W^\pm$ pair 
	\cite{Dittmar:1996ss, Dittmar:1996sp}. 
	As a result, the $\Delta R(\ell^\pm_1\ell^\pm_2)$ distribution peaks
	at a relatively lower value. On the other hand, same-sign leptons from all the 
	background events would have much wider separation as they originate directly 
	from $W^\pm/Z$ bosons that are either produced as a pair of primary objects or  
	are radiated from top/anti-top quark. Thus, an upper cut on the 
	$\Delta R(\ell_1^\pm \ell_2^\pm )(< 1.5)$ considerably enhances our signal to 
	background ratio. However, since the signal 
	distribution do not show a steeply falling nature, this cut would also reduce the number
	of detectable signal events.
	
A summary of all these cuts is as follows:
\begin{itemize}
\item ({\bf C1-1}): Our signal event is hadronically quiet, hence, we 
put a veto on any jet with $p_T > 30$ GeV. From Table~\ref{tab:3lmet_13},
it can be understood that the reduction in signal events is only the aftermath of branching 
ratio suppression for the gauge boson decay. On the contrary, all the SM backgrounds
with accompanying jets receive significant cutback in total cross section. 

\item ({\bf C1-2}): Next, to confirm the trilepton signature,  
we select at least three leptons with $ p_T > 10 $ GeV. This cut 
essentially removes most of the backgrounds arising from these SM processes:
$t+$ jets, $t \bar{t} +$ jets, $W^\pm+$ jets, $Z+$ jets, $W^{+}W^{-}+$ jets
with less number of isolated leptons, also evident from Table~\ref{tab:3lmet_13}.

\item ({\bf C1-3}): For further affirmation of trilepton signature, we reject any additional charged lepton 
with $p_T > 10 $ GeV. This, however, does not play a crucial role rather serves as a systematic
attestation.

\item ({\bf C1-4}): Furthermore, we claim that the same flavour opposite sign (SFOS) 
lepton invariant mass ${\rm M}_{\ell^+ \ell^-}$ should not lie between 
the window of 80-100 GeV to ensure that those are not directly produced from $Z$ boson.
This helps in suppressing the two most important background, i.e. $ZZ+\rm jets$ and $W^\pm Z+\rm jets$.

\item ({\bf C1-5}): For missing energy requirement, we demand 
$\mET > 30$ GeV. 

\item ({\bf C1-6}): The principal selection cut for
the same-sign dilepton has been imposed. For this, we demand $\Delta R (\ell^\pm_1 \ell^\pm_2 ) < $ 1.5. 
\end{itemize}

In Table~\ref{tab:3lmet_13}, we compile the effect of the aforementioned cuts for
both the signal (for all four benchmark points corresponding to both positive and 
negative scenario) and the SM background events and calculate the 
signal significance defined as 
\begin{eqnarray}
{\mathscr{S}} = \frac{N_S}{\sqrt{N_S + N_B}}\,,
\label{eq:signi}
\end{eqnarray}
where, $N_S(N_B)$ denotes the number of signal (background)
events at a specific luminosity.

\begin{table}[H]
	\centering
	\footnotesize
	\resizebox{17cm}{!}{
		\begin{tabular}{|p{1.7cm}|p{1.7cm}|c|c|c|c|c|c|p{1.7cm}|}
			\hline
			\multicolumn{2}{|c|}{}& \multicolumn{6}{|c|}{Effective cross section (fb) for background after the cut} & \\ \hline
			SM-background & Production Cross section (fb) 
			& C1--1 &~~ C1--2 ~~&~~ C1--3 ~~&~~ C1--4 ~~& C1--5 ~~& C1--6 & 
			\\ \hline 
			$t$+jets & $2.22 \times 10^5$ & 157.50 & 0 & 0 & 0 &  0 & 0 &\\ \hline 
			$t\bar{t}$+jets & $7.07 \times 10^5$  & 420.37 & 0 & 0 & 0 & 0 & 0 &\\ \hline
			$W^\pm$+jets & $1.54 \times 10^8$  & $4.96 \times 10^7$ & 0 & 0 & 0 & 0 & 0 &\\ \hline
			$Z$+jets & $4.54 \times 10^7$  & $1.37 \times 10^7$ & 0 & 0 & 0 & 0 & 0 &\\ \hline
			$W^+W^-$+jets & $8.22 \times 10^4$ & $4.76 \times 10^3$ & 0 & 0 & 0 & 0 & 0 &\\ \hline
			$ZZ$+jets & $1.10 \times 10^4$ & $6.17 \times 10^2$ & 10.05 & 5.77 & 0.08 & 0.04 & $\sim$ 0 &\\ \hline
			$W^\pm Z$+jets & $3.81 \times 10^4$ & $1.71 \times 10^3$ & 42.40 & 42.40 & 0.72 & 0.36 & 0.04 &\\\hline
			$W^+W^-Z$ & 83.10 & 1.17 & 0.09 & 0.07 & 0.01 &  $\sim$ 0 & 0 &\\ \hline
			$W^\pm ZZ$ & 26.80 & 0.39 & 0.03 & 0.03 & $\sim$ 0 &  0 & 0 &\\ \hline
			$t\bar{t}+W^\pm$ & 360 &  0.13 & 0.02 & $\sim$ 0 & 0 &  0 & 0 &\\ \hline
			$t\bar{t}+Z$ & 585 &  0.15 & 0.02 & 0.01 & $\sim$ 0 & 0 & 0 &\\ \hline
			$t\bar{t}+h$ & 400 &  0.02 & $\sim$ 0 & 0 & 0  &  0 & 0 &\\  \cline{1-8}
			\multicolumn{1}{|c|}{Total SM Background} & $2.005\times10^{8}$  & $6.33\times10^{7}$ & 52.60 & 48.30 & 0.81 & 0.40 & 0.04 &\\ \hline \hline
			\multicolumn{1}{|c|}{Positive scenario }&{Production Cross section (fb)} & \multicolumn{6}{|c|}{Effective cross section (fb) for signal after the cut} & {Luminosity (in $\rm fb^{-1}$) for 5$\sigma$ significance}\\ \hline
			\multicolumn{1}{|c|}{BP1} & 185.10 & 0.75 & 0.20 & 0.14 & 0.08 &  0.06 & 0.040 & 1250.0  \\ \hline 
			\multicolumn{1}{|c|}{BP2} & 158.70 & 0.65 & 0.16 & 0.11 & 0.06 &  0.05 & 0.034 & 1600.4 \\ \hline \hline
			\multicolumn{1}{|c|}{Negative scenario}&{Production Cross section (fb)}& \multicolumn{6}{|c|}{Effective cross section (fb) for signal after the cut} & {Luminosity (in $\rm fb^{-1}$) for 5$\sigma$ significance}\\ \hline
			\multicolumn{1}{|c|}{BP1} & 153.80 & 0.63 & 0.16 & 0.11 & 0.06 &  0.05 & 0.033 & 1675.8  \\ \hline 
			\multicolumn{1}{|c|}{BP2} & 134.70 & 0.55 & 0.15 & 0.10 & 0.05 &  0.04 & 0.030 & 1944.4  \\ \hline \hline
	\end{tabular}}
	\caption{\it Effective cross section obtained after each cut for both signal ($3\ell^\pm + \mET\,\,$) and background and the 
		respective required integrated luminosity for \rm 5$\sigma$ 
significance at $\rm 13~TeV$ LHC.}
 \label{tab:3lmet_13}
\end{table}

\begin{figure}[htbp!]
 \centering
\includegraphics[width=80mm,height=62mm]{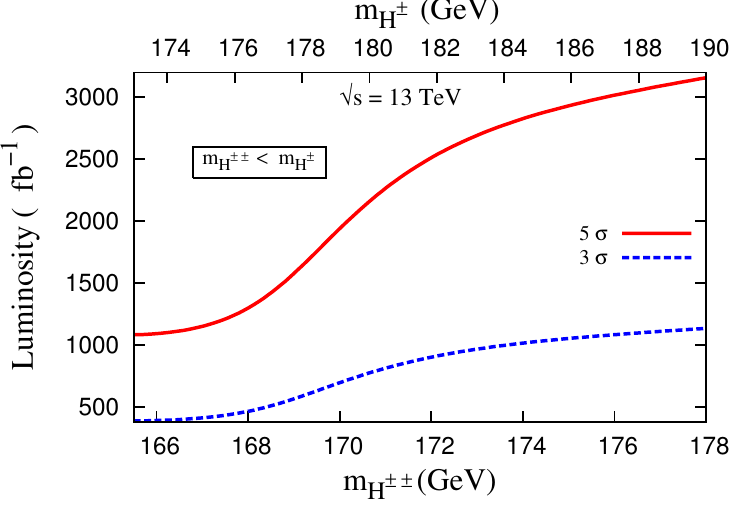} 
\includegraphics[width=80mm,height=62mm]{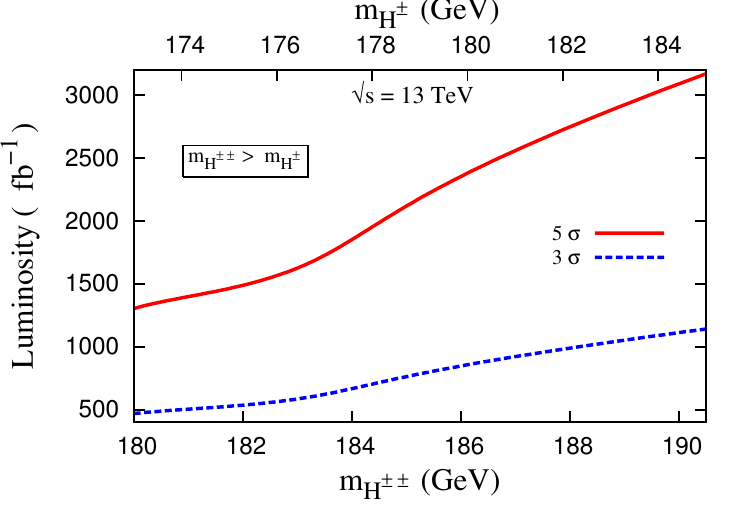}
\caption{\label{fig:sig} \textit{Left (Right) panel shows the required integrated luminosity for $3\ell^\pm + \mET \,$ 
final state with respect to charged Higgs masses at the LHC at $\sqrt{s}= \rm 13~TeV$ for positive (negative) scenario. 
The solid red coloured and blue coloured dashed curve correspond to constant signal significance at \rm 5$\sigma$ and \rm 3$\sigma$ respectively.}}

 \end{figure}

In Fig.\;\ref{fig:sig}, we show the required luminosity at 13 TeV LHC run
to reach signal significance $5\sigma$ (red solid curve) and $3\sigma$(blue dashed curve) respectively 
for a range of singly and doubly charged scalar mass for both the positive and negative 
scenarios. It is evident from the figure that 
a $5\sigma$ discovery reach at the 13 TeV LHC run can be achieved both in the
positive and negative scenario with a minimum integrated luminosity of $1250{~\rm fb}^{-1}$, 
even if the charged scalar masses lie at the ballpark of 170 GeV. 

For the negative scenario,
the required luminosity is a little higher due to the phase space suppression as the mass of the doubly
charged scalar is higher than that of the singly charged scalar. Therefore, the model prediction can only be probed through this channel with a
 High Luminosity LHC (HL-LHC) run.

Before concluding this section, we would like to comment that there can be two more 
signal topologies, namely: (a)$5\ell^\pm + \mET$ and (b) $3\ell^\pm + 4j + \mET \,$. However, the effective cross section
for channel (a) is suppressed by a factor of 3 due to the fact that $\frac{BR(Z \to \ell^+\ell^-)}{BR(Z \to \nu\nu)}
\sim \frac{1}{3}$. Moreover, due to the occurrence of two same-sign di-lepton, one arising from secondary decay of 
the doubly charged scalars
while the other from that of the singly charged scalar,  the angular separation cut will not be very helpful.
In addition to that, since the $Z$ boson decays to charged lepton final state, we can not impose cut C1-4 in this
case. We have checked that to attain $5\sigma$ signal significance we need at least $\sim 3100 ~{\rm fb^{-1}}$ luminosity, which
can only be probed with very high luminosity run of the LHC. \\
Coming to the second topology, i.e. the $3\ell^\pm + 4j + \mET \,$ channel, the effective cross-section is 4 times larger
than the $3\ell^\pm + \mET \,$ channel. In spite of that larger cross-section the significance is much lesser due to the fact the
$W^\pm Z+~{\rm jets}$ SM background mimics the signal and also the angular separation cut is not effective as mentioned before.
As a result, we require minimum $2350 ~{\rm fb^{-1}}$ luminosity to reach  $5\sigma$ significance.


\subsubsection{$2\ell^\pm + 4j + \mET$ }

This particular channel receives contributions 
from both doubly charged pair production $(H^{++} H^{--})$ and singly
charged scalar associated with doubly charged scalar production $(H^{\pm\pm}H^{\mp})$  processes.  
The decay chains which lead to this final state
are depicted by:
\begin{subequations}
\begin{eqnarray}
p p \to H^{++} H^{--} &\to & (W^+ W^+) + (W^-W^-) \to (\ell^+ \ell^+) + 4j 
+ \mET \,\label{pair}~,\\
p p \to H^{++} H^- &\to & (W^+ W^+) + (W^- Z) \to (\ell^+ \ell^+) + 4j + \mET~.
 \label{asso}
\end{eqnarray}
\end{subequations}

The other charge combination should also be included. The LSD nature of this final state 
makes it a very promising channel to explore the Type II seesaw scenario 
at the LHC. In addition to the LSD, the signal channel also contains four 
hard jets and a moderate amount of missing transverse energy.

Before we embark our study of selection cuts on different 
kinematic variables associated with this signal channel, we would like to 
discuss the general features of this final state. The LSDs originate from the 
decays of two same-signed $W^\pm$ bosons. Here, we expect that the transverse 
momentum $(p_T)$ spectrum of these leptons would be very similar to those 
charged leptons of our previous channel simply because of their identical 
origin, the $W^\pm$ boson. The neutrinos produced in association with the charged
leptons from the decay of $W^\pm$ bosons are the main source of the observed 
$\mET$. However, there is a small contribution to it which comes from the uncertainty 
associated with jet energy measurement.
We expect that the $\mET$ ~distribution would not be very different from 
that of the previous channel, as displayed in the 
left panel of Fig.\;\ref{fig:met}.

\begin{figure}[ht!]
\centering
\includegraphics[width=8cm,height=5cm, angle=0]{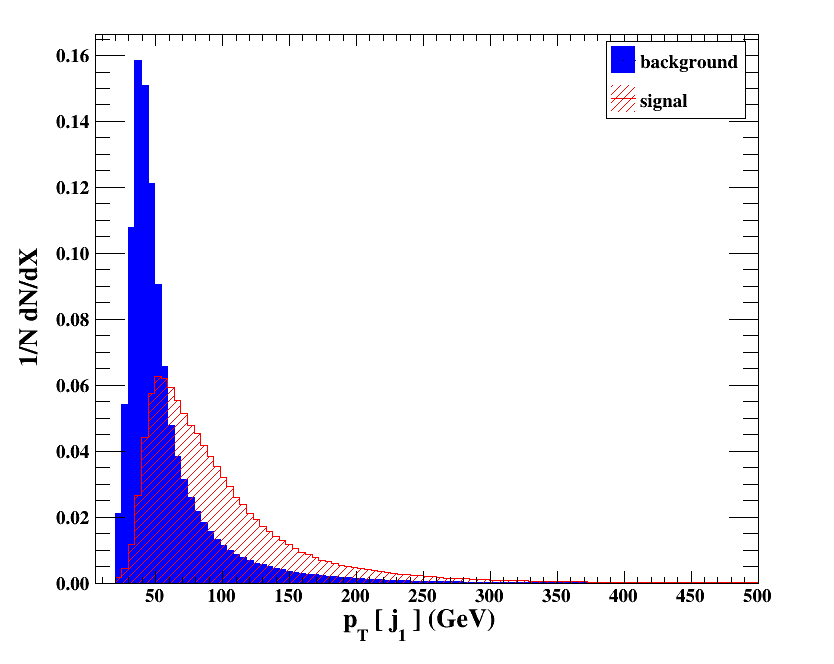} 
\includegraphics[width=8cm,height=5cm, angle=0]{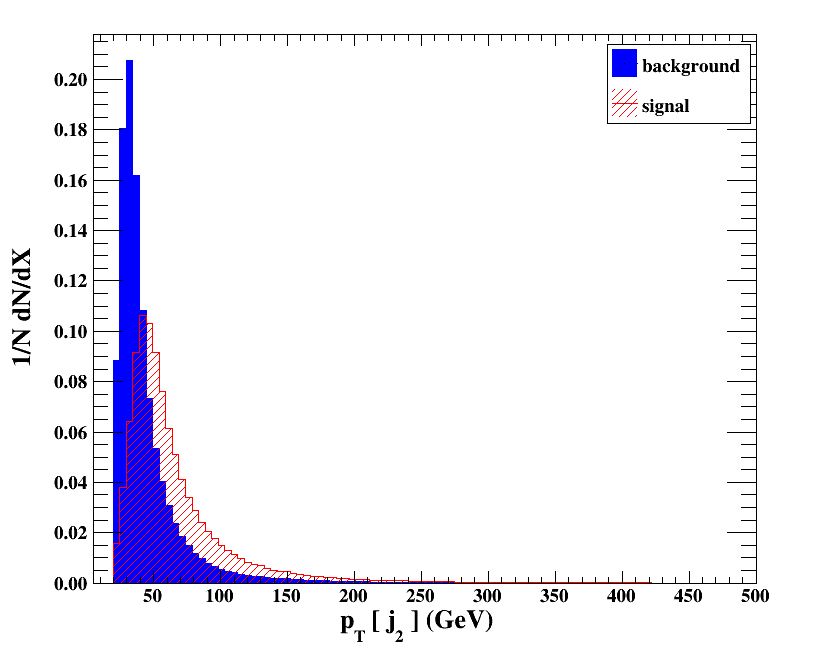} \\
\includegraphics[width=8cm,height=5cm, angle=0]{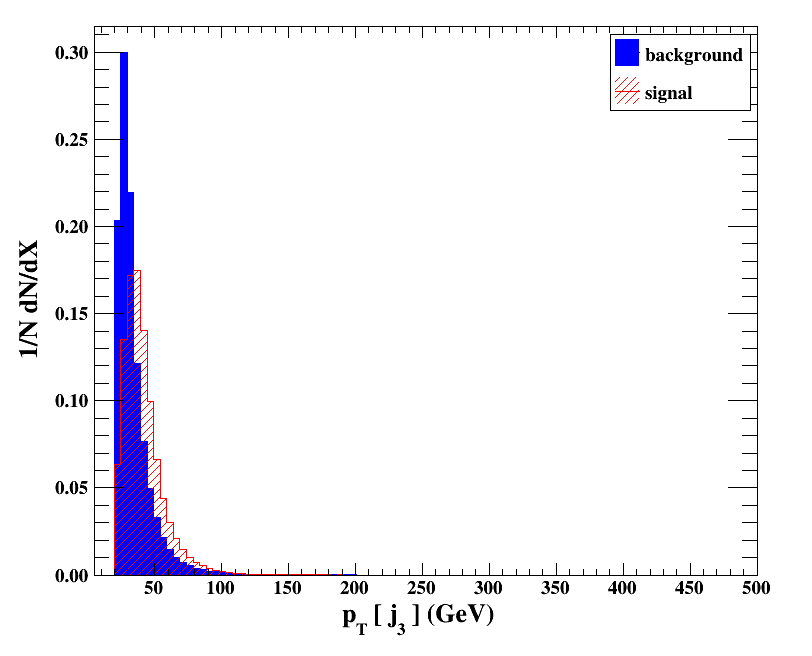} 
\includegraphics[width=8cm,height=5cm, angle=0]{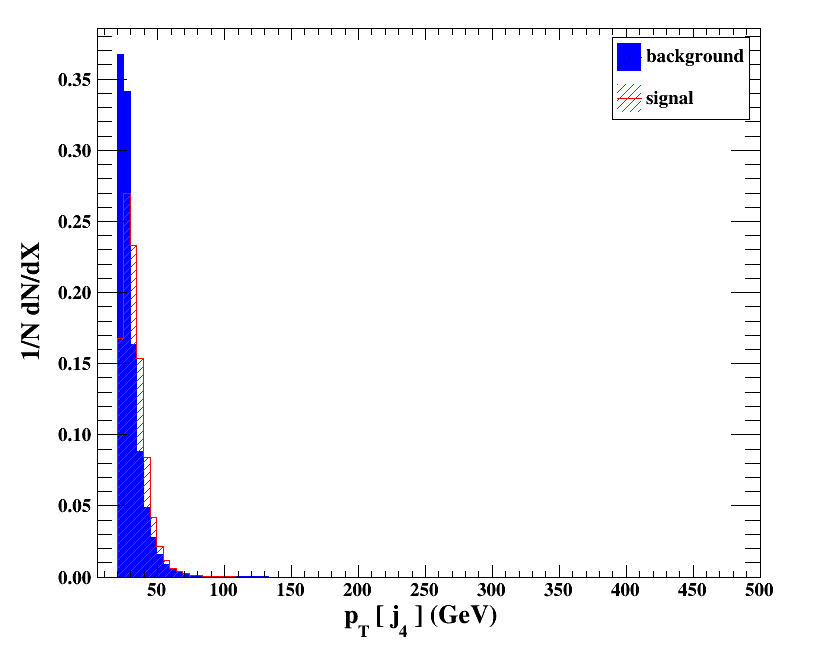}
 \caption{\label{fig:jet_pT} \textit{Transverse momentum $(p_T)$ distribution (normalized) of the four leading jets 
for the final state $2\ell^\pm + 4j + \mET$ for the benchmark BP1 of positive scenario. 
}}
 \end{figure}

In Fig.\;\ref{fig:jet_pT}, we show the $p_T$ distributions of the four hardest 
jets for both the signal and the background events. 
By looking at the shape of the jet 
$p_T$ distributions, it is very clear that the signal jets are relatively 
harder than that of the SM backgrounds. This can attributed to the fact that 
for the signal process, these jets originate from the decays of different 
$W^\pm$ and $Z$ bosons which in turn produced with some transverse momentum 
from the decay of heavy $H^{\pm}$ and $H^{\pm\pm}$ Higgs bosons. 
On the other hand significant fraction of the SM background jets 
originate from the initial state of final state radiation, making them 
relatively softer compared to the signal jets. So, we demand that our signal events should have four jets starting from a leading jet 
with $p_T(j_1) > 60$ GeV and subsequently, the next to leading jet with $p_T(j_2) > 40$ GeV while the
rest of the jets with at least $p_T > 20$ GeV are asked.
However, these jets do not originate from $b$-quark hadronisation 
and so a veto on $b$-tagged jet is extremely useful in increasing
the signal significance.

Akin to our previous case, the distribution of the angular separation $\Delta R(\ell^\pm\ell^\pm)$
	between the two same-sign leptons follows the same nature as shown in Fig.~\ref{fig:dr}. However, in this signal
	channel absence of any additional charged lepton makes the cut more severe to the
	background with respect to the signal. We will see that an upper cut on the 
	$\Delta R(\ell_1^\pm \ell_2^\pm )(< 1.5)$ helps gaining a huge signal significance. 
	
Following the general features of the distributions of different 
kinematical variables, we now implement our selection criteria :
\begin{itemize}
\item ({\bf C2-1}): As explained, our signal is exempted from any $b$-jets, hence 
we can safely reject events with $b$-tagged jets of $p_T(b) > 40$ GeV. This cut 
drastically reduces the SM background events arising from 
$t\bar t+ \rm jets $, $t+\rm jets$ and $t\bar{t}$+$(h/W^\pm/Z)$ processes.  
\item ({\bf C2-2}): To guarantee that only 4 jets are present in the
events, we reject any additional jets with $p_T(j_5) > 20$ GeV.   
\item ({\bf C2-3}): Our signal also contains two isolated charged lepton
and thus a veto on any additional leptons with $p_T > 10$ GeV
is applied. 
\item ({\bf C2-4}): Now, from the jet distribution, 
we choose the $p_T$ of the leading jet to be at least greater than $p_T(j_1) > 60$ GeV.
This helps in a modest drop in background events.
\item ({\bf C2-5}): Moreover, for the next sub-leading jet 
we demand $p_T(j_2) > 40$ GeV to further subdue the background events. 
\item ({\bf C2-6}): A decent selection cut on the missing transverse energy is then
applied as $\mET > 30$ GeV. 
\item ({\bf C2-7}): At the end,  we demand $\Delta R (\ell^\pm_1 \ell^\pm_2 ) < $ 1.5
	as the most effective cut. We find that this cut is crucial for achieving higher signal significance. 
\end{itemize}

We list the gradual effect of the selection cuts on the
signal and background events in Table~\ref{2l4j_13}.
We estimate the required integrated luminosity (using the Eq.~(\ref{eq:signi}))
to observe a $5\sigma $ significance 
corresponding to each benchmark points for both the scenarios at the 13 TeV LHC.
It is evident from Table~\ref{2l4j_13} that this signal can even be observed with
a $5\sigma $ significance with only a mere $27~\rm fb^{-1}$ luminosity 
for the benchmark point of highest cross-section. 
For the rest of the benchmark also the required luminosity is quite low
(30-50 $\rm fb^{-1}$). 
Hence, we may expect to see this final state at 
$5 \sigma$ signal significance at the current run of the 13 TeV LHC.

\begin{table}[htbp!]
	\centering
	\footnotesize
\resizebox{17cm}{!}{
	\begin{tabular}{|p{1.7cm}|p{2.1cm}|c|c|c|c|c|c|c|p{1.7cm}|}
		\hline
		\multicolumn{2}{|c|}{}& \multicolumn{7}{|c|}{Effective cross section (fb) after the cut} & \\ \hline
		SM-background & Production Cross section (fb) 
		& C2--1 & C2--2 & C2--3 & C2--4 & C2--5 & C2--6 & C2--7 & 
		  \\ \hline 
		$t$+jets & $2.22 \times 10^5$ & $8.46 \times 10^4$ & $8.01\times 10^4$ & $8.01\times 10^4$ & $4.89\times 10^4$ & $3.44\times 10^4$ & $1.54\times 10^4$  & 0 &\\ \hline 
		$t\bar{t}$+jets & $7.07 \times 10^5$ & $1.58 \times 10^5$ & $1.23 \times 10^5$ & $1.23 \times 10^5$ & $9.92 \times 10^4$ & $8.15 \times 10^4$  & $5.58 \times 10^4$  & 0 &\\ \hline
		$W^\pm$+jets & $1.54 \times 10^8$ & $1.52 \times 10^8$ & $1.52\times 10^8$ & $1.52\times 10^8$ & $1.24\times 10^7$ & $8.17\times 10^6$ & $1.75\times 10^6$  & 0 &\\\hline
		$Z$+jets & $4.54 \times 10^7$ & $4.27 \times 10^7$ & $4.27\times 10^7$ & $4.27\times 10^7$ & $3.76\times 10^6$ & $2.48\times 10^6$ & $4.65\times 10^5$  & 0 &\\ \hline
		$W^+W^-$+jets & $8.22 \times 10^4$ & $7.84 \times 10^4$ & $7.55\times 10^4$ & $7.55\times 10^4$ & $3.48\times 10^4$ & $2.39\times 10^4$ &  $1.04\times 10^4$ & 0 &\\ \hline
		$ZZ$+jets & $1.10 \times 10^4$ & $8.96 \times 10^3$ & $8.67 \times 10^3$ & $8.65 \times 10^3$ & $4.27 \times 10^3$ & $2.89 \times 10^3$ & $1.16 \times 10^3$ &  0.05 &\\ \hline
		$W^\pm Z$+jets & $3.81 \times 10^4$ & $3.33 \times 10^4$ & $3.13\times 10^4$ & $3.12\times 10^4$ & $1.67\times 10^4$ & $1.18\times 10^4$  & $5.76\times 10^3$ & 1.68 &\\\hline
                $t\bar{t}+W^\pm$ & 360 & 78.00 & 55.15 & 55.00 & 47.00 & 39.80 & 33.00   & 0.13 &\\ \hline
		$t\bar{t}+Z$ & 585 & 110.00 & 68.20 & 67.00 & 59.60 & 52.04 & 44.30   & 0.04 &\\ \hline
		$t\bar{t}+h$ & 400 & 46.00 & 27.40 & 27.20 & 24.50 & 21.65 & 18.30    & 0.04 &\\  \cline{1-10}
		\multicolumn{1}{|c|}{Total SM Background} & $2.005\times10^{8}$ & $1.95\times10^{8}$ & $1.94\times10^{8}$ & $1.94\times10^{8}$ & $1.64\times10^{7}$ & $1.08\times10^{7}$ & $2.31\times10^{6}$  & 1.94 &  \\ \hline  \hline
	\multicolumn{1}{|c|}{Positive scenario}&{Production Cross section (fb)}& \multicolumn{7}{|c|}{Effective cross section (fb) for signal after the cut} & {Luminosity (in $\rm fb^{-1}$) for 5$\sigma$ significance}\\ \hline
	\multicolumn{1}{|c|}{BP1} & 311.40 & 253.00 & 210.70 & 206.70 & 181.72 & 158.90 & 126.24 & 1.90 & 26.6 \\ \hline 
	\multicolumn{1}{|c|}{BP2} & 259.30 & 211.23 & 175.92 & 172.64 & 151.75 & 132.66 & 105.42 & 1.55  & 36.3\\ \hline \hline 
        \multicolumn{1}{|c|}{Negative scenario}&{Production Cross section (fb)}& \multicolumn{7}{|c|}{Effective cross section (fb) for signal after the cut} & {Luminosity (in $\rm fb^{-1}$) for 5$\sigma$ significance}\\ \hline
        \multicolumn{1}{|c|}{BP1} & 246.23 & 200.00 & 166.50 & 163.32 & 143.61 & 125.54 & 99.76 & 1.50   & 38.2 \\ \hline 
	\multicolumn{1}{|c|}{BP2} & 219.30 & 177.74 & 148.03 & 145.22 & 127.70 & 111.63 & 88.71 & 1.30  & 47.9\\ \hline \hline
	\end{tabular}}
	\caption{\it Effective cross section obtained after each cut for both signal ($2\ell^\pm + 4j + \mET$) and background and the 
   	respective required integrated luminosity for \rm 5$\sigma$ significance at $\rm 13~TeV$ LHC.}
	\label{2l4j_13}
\end{table}

\begin{figure}[htbp!]
\centering
\includegraphics[width=80mm,height=62mm]{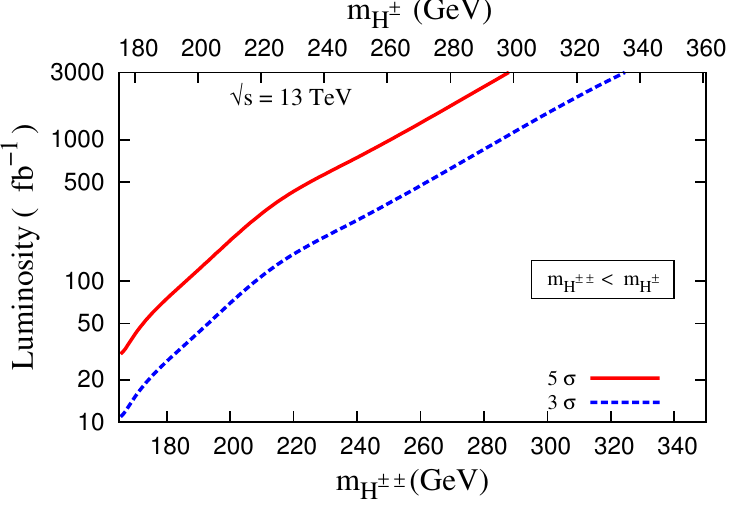} 
\includegraphics[width=80mm,height=62mm]{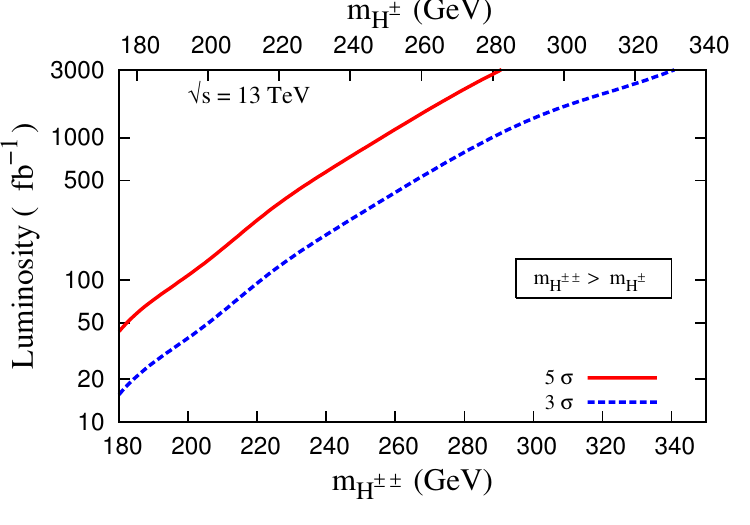}
\caption{\label{fig:lumi} \textit{Left (Right) panel shows the required integrated luminosity for $2\ell^\pm + 4j + \mET$ final state with respect to charged Higgs masses at the LHC at $\sqrt{s}=\rm 13~TeV$ for positive (negative) scenario. The solid red coloured and blue coloured dashed curve correspond to constant signal significance at \rm 5$\sigma$ and \rm 3$\sigma$ respectively.}}
\end{figure}

After obtaining a very encouraging signal to background ratio for this 
channel, we now try to get an estimate of the singly and doubly charged Higgs
mass range which the 13 TeV LHC can probe and the integrated 
luminosity required for this purpose. In Fig.~\ref{fig:lumi} we show the
$3\sigma $ and $5\sigma$ significance contour in the mass of heavy charged 
scalar ($m_{H^{\pm \pm}}$ (lower axis) and $m_{H^\pm}$ (upper axis)) 
and integrated luminosity plane for positive (left) and negative (right) 
scenarios respectively. One can see that running at 13 TeV energy and with the 3 ${\rm ab}^{-1}$ 
integrated luminosity the LHC can probe the singly(doubly) charged scalar 
masses up to 300(280) GeV and up to 335(320) GeV with
5$\sigma$ and 3$\sigma$ significance for the positive scenario 
(left panel of Fig.~\ref{fig:lumi}). For the negative scenario (right panel of Fig.~\ref{fig:lumi}), 
the LHC reach is almost identical with the positive scenario 
except with a reverse mass hierarchy between the singly and doubly charged scalars.
It is also very clear that for the masses around 200 GeV, the integrated luminosity of about a 100~$\rm fb^{-1}$
is more than sufficient to get the discovery reach of 5$\sigma$.
We thus strongly motivate this channel to be looked at the current run of the LHC.

In passing, we would like to comment that the channel $(\ell^+\ell^-) + 4j + \mET$
has same effective cross section as  our second signal process. But
absence of a pair of same sign lepton plague the detection of this channel.
Even with $3~{{\rm ab}^{-1}}$ luminosity, only $1\sigma$ significance
can be achieved.

Before making our conclusion, we would like to stress 
that while estimating signal significances for both the channels
we have not taken into account
different the experimental issues arising from the electron charge 
misidentification, jet faking as leptons and photon conversions into 
lepton pair. However, we have found that the misidentification 
probability of a jet to be an isolated electron is around $(0.1-0.2)\%$ 
for $p_T \simeq 30 ~\rm GeV$ for the {\bf Tight} 
criterion~\cite{ATLAS:2016iqc}. This would imply that 
for the $3\ell +\mET~~{\rm and}~2\ell + 4j +\mET ~\rm$ signal channels, 
all the SM multi-jet background processes will be down by 
order $10^{-9}$ and $10^{-6}$ respectively. The electron misidentification 
probability is also very small $({\cal O}(0.1\%))$ in the central rapidity 
region~\cite{Khachatryan:2015hwa}. As far as the photon conversions 
into $e^+e^-$ probability is concerned, we expect it to be of similar 
level as other two aforementioned fake rates.
One should note that our search result is
independent of the mass hierarchy of the charged scalars and appear to
be equally discernible at the LHC. Also, we strongly motivate the
same-sign dilepton plus jets and missing energy channel as an aspiring
search channel at the current LHC run.

\section{Conclusion}

With an intention to solve the infamous vacuum instability problem, we have recalled the Type-II seesaw
model where the SM scalar sector is extended with an $SU(2)_L$ triplet scalar. 
We have found that the additional scalar fields can certainly surmount the 
instability problem and provide us with an absolutely stable vacuum even up 
to the Planck scale. The stipulation of the stability, unitarity and T-parameter 
constraints altogether up to the 
high cut-off scale severely affect the allowed model parameter space favored 
at the EW scale. To exemplify, we have chosen two distinct values of the 
triplet vev $v_t$ (1 and 3 GeV) and show the amount
of parameter space spared by the above theoretical constraints. We have observed that the
requirement of an absolutely stable vacuum up to the high Planck scale
have pushed the neutral scalar 
mixing angle $(\alpha)$ to a quite small range of value
for the non-degenerate mass scenario $(m_H(m_A) \gg m_h)$
and peaks around a small positive value for considerably large $m_H$. 
The peak shifted to a more positive value with increasing $v_t$. 
On the other hand, the mass differences
among all the other non-standard scalar masses have narrowed down to small values. 
Besides an almost degenerate non-standard CP-even ($H$) and CP-odd ($A$) neutral scalar,
the doubly and singly charged scalars also become closer in mass. 

Next, we have inspected the expectancy to detect the model prediction, 
available from high-scale valid scenario, at the current and high 
luminosity run of the LHC. We have particularly investigated the associated 
production of the singly and doubly charged scalars. 
Depending on the mass hierarchy, two possible scenarios (positive and negative)
exist which however, at the end, yielded similar signal significance. In the allowed parameter space, with
appreciable production cross section, the masses of the charged scalar can 
presumably be chosen around 200 GeV. This mass choice along with
our choice of triplet vev, fix the possible decay modes of the charged scalars.
Correspondingly it give rises to two specific final states at the collider,
($3\ell^\pm +  \mET \,$) and  $(2\ell^\pm + 4j + \mET)$ at the 13 TeV LHC run. 
A proper cut-based analysis with detector simulation reveals that the first 
channel can only be probed at the 13 TeV LHC with high integrated luminosity
around 1250 $\rm fb^{-1}$ and should be promoted for a HL-LHC. 
On the other hand, the second signal channel has better observability at the 
13 TeV LHC due to larger production rate and better cut efficiencies. We have found that even a $5\sigma$ discovery reach is possible with the present LHC 
data with only around $40~\rm fb^{-1}$ luminosity. 
We would like to mention that our estimation of the signal
significance is based on
simple cut based analysis, where we did not include some experimental issues
arising from jet faking as leptons, lepton charge misidentification and
photon conversions into lepton pairs. We have argued semi-quantitatively that
the first two fake rates would not affect our main conclusions and expect
that the photon conversions into lepton pairs would also be rather small.
Any quantitative statements regarding aforementioned fake rates are beyond the
scope of this phenomenological analysis. We agree that the proper inclusion
of these fake rates as well as systematic uncertainties in the SM background 
estimation would certainly modify the signal significance quoted in this
paper, hence our conclusions may be considered as an indicative one. 
We hope that the results presented in this paper would encourage the 
experimental collaborations to perform more dedicated analysis in this 
direction to probe this scenario both in the current and future run of the LHC.

\section{Acknowledgement}
N.G. and A.S. would like to thank Subhadeep Mondal for useful discussions.
N.G. would like to acknowledge the Council of Scientific and Industrial Research (CSIR), Government of India for financial
support. We all are thankful to Subir Sarkar and Manas Maity for useful discussions.

\appendix
\section*{Appendix}
\section{One loop RG equations} 
\label{RGE_all}

Here, we will present the one loop RGEs of all the relevant couplings (gauge, Yukawa and scalar quartic couplings) of the Type-II seesaw model \cite{Chao:2006ye,Schmidt:2007nq}. For convenience, we introduce the shorthand notation 
${\cal D}\equiv 16\pi^2\frac{\rm d}{\rm {d(ln\mu)}}$.

\paragraph{Gauge and top Yukawa couplings:}
The RGE for the gauge couplings, 
\begin{subequations}
	\begin{eqnarray}
	{\cal D}g_1 &=& \frac{47}{10}{g_1}^3\,,  \label{gp_rg} \\
	{\cal D}g_2 &=& -\frac{5}{2}g_2^3\,,   \label{g_rg}\\
	{\cal D}g_3 &=& -7g_3^3\,.  \label{gs_rg} 
	\end{eqnarray}
	The RGE for the top Yukawa coupling,
	\begin{eqnarray}
	{\cal D}y_t &=& y_t\left( \frac{9}{2}y_t^2 -\left(8g_3^2 + \frac{9}{4}g_2^2 + \frac{17}{20}{g_1}^2\right)\right)\,,  \label{top_rg_t1}
	\end{eqnarray}
	\label{gauge_top_rg}
\end{subequations}

where, $g_1 = \sqrt{\frac{5}{3}}g^\prime$ with GUT renormalization.
The boundary values of the gauge couplings at $M_Z$ are taken to be~\cite{Olive:2016xmw}
	\begin{eqnarray}
	g_1(M_Z) &=& \sqrt{\frac{5}{3}} 0.3569 \,, \qquad
	g_2(M_Z) = 0.6531 \,, \qquad
	g_3(M_Z) = \sqrt{4 \pi 0.1185} \,.
	\end{eqnarray}
 While the boundary condition for the top Yukawa interaction at the electroweak scale is fixed by
  $y_t(M_t) = \frac{\sqrt{2}m_t(M_t)}{v}$. For the matching condition between the
  $\overline{MS}$ mass of top quark to its pole mass, we consider 
  the dominant one-loop QCD correction as  $m_t(M_t)\sim M_t\left( 1 - \frac{4}{3\pi} \alpha_s(M_t)\right)$, $M_t$ being the
top pole mass fixed at 173 GeV throughout the analysis and the strong coupling $\alpha_s(M_t)$ is derived 
from the gauge coupling running from $M_Z$ to $M_t$. 

\paragraph{Scalar quartic couplings:}
We express the RGEs of the scalar quartic coupling with a 
redefinition of the coupling to match with the potential notation
of Ref.~\cite{Schmidt:2007nq} which can be translated from our notation in the following way.
\begin{subequations}
\begin{eqnarray}
\Lambda_0 &=& \frac{\lambda}{2} \,, \\
\Lambda_1 &=& 2 \lambda_2 + 2 \lambda_3 \,, \\ 
\Lambda_2 &=& -2 \lambda_3 \,, \\ 
\Lambda_4 &=& \lambda_1 + \frac{\lambda_4}{2} \,, \\
\Lambda_5 &=& -\frac{\lambda_4}{2}\,.
\end{eqnarray}
\label{eq:translation}
\end{subequations}

The RGEs for the five quartic couplings that appear are then given by,
\begin{eqnarray}
{\cal D}\Lambda_i = \beta_{\Lambda_i} + G_i \,, (i=0,1,2,4,5)\,,
\label{lambda_rg}
\end{eqnarray}
where, $\beta_{\Lambda_i}$ and $G_i$ are as follows:
\begin{subequations}
\begin{eqnarray}
\beta_{\Lambda_0} &=& 12 \Lambda_0^2 + 6 \Lambda_4^2 + 4 \Lambda_5^2 \,, \label{l0_rg_l}\\
\beta_{\Lambda_1} &=& 14 \Lambda_1^2 + 4 \Lambda_1 \Lambda_2 +  2 \Lambda_2^2 + 4 \Lambda_4^2 + 4 \Lambda_5^2\,, \label{l1_rg_l}\\
\beta_{\Lambda_2} &=& 3 \Lambda_2^2 + 12 \Lambda_1 \Lambda_2 - 8 \Lambda_5^2 \,, \label{l2_rg_l}\\
\beta_{\Lambda_4} &=& \Lambda_4 \left(8 \Lambda_1 + 2 \Lambda_2 + 6 \Lambda_0 + 4 \Lambda_4 +8 \Lambda_5^2 \right)\,, \label{l4_rg_l}\\
\beta_{\Lambda_5} &=& \Lambda_5\left(2 \Lambda_1 - 2 \Lambda_2 + 2 \Lambda_0 + 8 \Lambda_4\right)\,, \label{l5_rg_l} 
\end{eqnarray}
\label{lambda_rg_l}
\end{subequations}
and,
\begin{subequations}
\begin{eqnarray}
G_0 &=& \left(12  y_t^2-\left(\frac{9}{5} g_1^2 + 9 g_2^2\right)\right) \Lambda_0 + \frac{9}{4} \left(\frac{3}{25} g_1^4 + \frac{2}{5} g_1^2 g_2^2 + g_2^4\right) - 12  y_t^4\,, \label{l0_rg_g}\\
G_1 &=& -\left(\frac{36}{5} g_1^2 + 24  g_2^2\right) \Lambda_1 + \frac{108}{25}  g_1^4 + 18  g_2^4 + \frac{72}{5} g_1^2 g_2^2 \,, \label{l1_rg_g}\\
G_2 &=& -\left(\frac{36}{5} g_1^2 + 24  g_2^2\right) \Lambda_2 + 12 g_2^4 -  \frac{144}{5} g_1^2 g_2^2 \,, \label{l2_rg_g}\\
G_4 &=& \left(6 y_t^2 - \left(\frac{9}{2} g_1^2 + \frac{33}{2} g_2^2\right)\right)\Lambda_4+ \frac{27}{25} g_1^4 + 6 g_2^4  \,, \label{l4_rg_g}\\
G_5 &=& \left(6 y_t^2 - \left(\frac{9}{2} g_1^2 + \frac{33}{2} g_2^2\right)\right)\Lambda_5- \frac{18}{5} g_1^2 g_2^2\,.  \label{l5_rg_g}
\end{eqnarray}
\label{lambda_rg_g}
\end{subequations}



\end{document}